\documentclass{emulateapj}
\usepackage{graphicx,mathptm}

\newcommand{\suz}{{\it Suzaku}}
\newcommand{\xmm}{{\it XMM-Newton}}

\shorttitle{Fe K emission lines in neutron star LMXBs} 
\shortauthors{Cackett et al.}

\begin{document}

\title{Relativistic Lines and Reflection from the Inner Accretion Disks Around Neutron Stars}

\author{Edward M. Cackett \altaffilmark{1,8}}
\author{Jon M. Miller\altaffilmark{1}}
\author{David R. Ballantyne\altaffilmark{2}}
\author{Didier Barret\altaffilmark{3}}
\author{Sudip Bhattacharyya\altaffilmark{4}}
\author{Martin Boutelier\altaffilmark{3}}
\author{M. Coleman Miller\altaffilmark{5}}
\author{Tod E. Strohmayer\altaffilmark{6}}
\author{Rudy Wijnands\altaffilmark{7}}

\email{ecackett@umich.edu}

\affil{\altaffilmark{1}Department of Astronomy, University of Michigan, Ann Arbor, MI, 48109, USA}
\affil{\altaffilmark{2}Center for Relativistic Astrophysics, School of Physics, Georgia Institute of Technology, Atlanta, GA 30332, USA}
\affil{\altaffilmark{3}Centre d' Etude Spatiale des Rayonnements, CNRS/UPS, 9 Avenue du Colonel Roche, 31028 Toulouse Cedex04, France}
\affil{\altaffilmark{4}Department of Astronomy and Astrophysics, Tata Institute of Fundamental Research, Mumbai 400005, India}
\affil{\altaffilmark{5}Department of Astronomy, University of Maryland, College Park, MD 20742-2421, USA}
\affil{\altaffilmark{6}Astrophysics Science Division, NASA/GSFC, Greenbelt, MD 20771, USA}
\affil{\altaffilmark{7}Astronomical Institute `Anton Pannekoek', University of Amsterdam, Science Park 904, 1098 XH, Amsterdam, The Netherlands}

\altaffiltext{8}{Chandra Fellow}

\begin{abstract}

A number of neutron star low-mass X-ray binaries have recently been discovered to show broad, asymmetric Fe K emission lines in their X-ray spectra.  These lines are generally thought to be the most prominent part of a reflection spectrum, originating in the inner part of the accretion disk where strong relativistic effects can broaden emission lines.  We present a comprehensive, systematic analysis of \suz{} and \xmm{} spectra of 10 neutron star low-mass X-ray binaries, all of which display broad Fe K emission lines.  Of the 10 sources, 4 are Z sources, 4 are atolls and 2 are accreting millisecond X-ray pulsars (also atolls). The Fe K lines are well fit by a relativistic line model for a Schwarzschild metric, and imply a narrow range of inner disk radii (6 -- 15 $GM/c^2$) in most cases.  This implies that the accretion disk extends close to the neutron star surface over a range of luminosities.  Continuum modeling shows that for the majority of observations, a blackbody component (plausibly associated with the boundary layer) dominates the X-ray emission from 8 -- 20 keV. Thus it appears likely that this spectral component produces the majority of the ionizing flux that illuminates the accretion disk.  Therefore, we also fit the spectra with a blurred reflection model, wherein a blackbody component illuminates the disk.  This model fits well in most cases, supporting the idea that the boundary layer is illuminating a geometrically thin disk. 

\end{abstract}

\section{Introduction}

X-ray emission lines from the innermost accretion disk are well-known
in supermassive and stellar-mass black holes, where they are
shaped by relativistic effects \citep[for a review, see][]{miller07}.  The utility of such lines is two-fold: they can be used to
constrain the spin of the black hole, and they can be used to
constrain the nature of the innermost accretion flow, particularly the
proximity of the disk to the black hole \citep{brenneman06,miniutti09,reis09_spin,miller06,miller08,miller09,reynolds09,wilkinson09,schmoll09,fabian09}.

Disk lines are produced in a fairly simple manner:  a source of hard X-ray emission that is external to the disk is all that is required.
The specific nature of the hard X-ray emission -- whether
thermal or non-thermal, whether due to Compton-upscattering
\citep[e.g.][]{gierlinski99} or emission from the base of a jet \citep{markoff04,markoff05}, or even a hot blackbody
-- is less important than the simple fact of a continuum source with substantial ionizing flux. The most prominent line produced in this process is typically an Fe K line, due to its abundance and fluorescent yield, however other lines can also
be produced.  The overall interaction is known as ``disk reflection"
and more subtle spectral features, including a ``reflection
hump" peaking between 20 -- 30~keV, are also expected \citep[see, e.g.][]{george91,magdziarz95,nayakshin01,ballantyne01,ross07}.

Largely owing to a combination of improved instrumentation and
concerted observational efforts within the last two years,
asymmetric Fe K disk lines have been observed in 8 neutron star
low-mass X-ray binaries \citep{bhattacharyya07,cackett08, pandel08,
dai09, cackett_j1808_09, papitto09, shaposhnikov09, reis09_1705,
disalvo09,iaria09}.  In the case of neutron stars, disk lines can be
used in much the same way as in black holes to determine the inner disk radius.  The radius of a
neutron star is critical to understanding its equation of state; disk
lines set an upper limit since the stellar surface (if not also a
boundary layer) truncates the disk.  In the case of neutron stars
harboring pulsars, disk lines can be used to trace the radial extent
of the disk and to obtain a magnetic field constraint \citep{cackett_j1808_09}.  Finally, in a number of neutron stars, hot quasi-blackbody emission (potentially from the  boundary layer) provides most of the flux required to ionize iron; the fact that the disk intercepts this flux suggests that the inner disk is geometrically thin, or at least thinner than the vertical extent of the boundary layer.

Disk line spectroscopy provides an independent view on the inner
accretion flow in neutron star LMXBs, wherein some constraints have
been derived using X-ray timing \citep[for a review, see][]{vanderklis06}.
The higher frequency part of so-called kHz QPO pairings may reflect
the Keplerian orbital frequency in the inner disk \citep[e.g.][]{stella98},
for instance.  Trends in the coherence of the low
frequency part of kHz QPO pairs may indicate the point at which the
disk has reached an inner stable orbit \citep{barret06}, though see \citet{mendez06} for an opposing view.
Thus, X-ray timing and spectroscopy
offer independent windows on the neutron star and inner accretion flow
in LMXBs.  It has been noted that with the assumption of Keplerian frequencies, combining the upper kHz QPO frequency with the radius given by a line measurement yields a mass determination  \citep{piraino00,cackett08}.

In order to understand possible differences in the innermost accretion
flow onto neutron star low-mass X-ray binaries, and in order to derive
joint constraints on fundamental neutron star parameters, it is
important to analyze the full sample of disk lines using a consistent,
systematic approach.  Herein, we analyze archival data
from \suz{} \citep{mitsuda07} and \xmm{} \citep{jansen01},
using consistent and systematic reduction and
modeling procedures.  Each spectrum is fit both with phenomenological
models, and with physically-motivated disk
reflection models.  In addition to Fe K lines reported previously, we
also present the discovery of broad lines in two additional sources,
GX~17+2 and HETE~J1900.1$-$2455.  The total number of neutron star LMXBs with
relativistic lines now stands at 10.

\section{Sample of Neutron Stars}
\label{sample}

In this paper we analyze data from 10 neutron star LMXBs, eight of which have recently had detections of broad Fe K emission lines \citep{bhattacharyya07,cackett08, pandel08,
dai09, cackett_j1808_09, papitto09, shaposhnikov09, reis09_1705,
disalvo09,iaria09}.  In addition, we also present evidence for relativistic Fe K emission lines in two further sources (GX~17+2 and HETE~J1900.1$-$2455).  The total sample consists of a variety of different types of neutron star LMXBs: four atolls (Ser~X-1, 4U~1636$-$53, 4U~1705$-$44, 4U~1820$-$30), four Z sources (GX~17+2, GX~340+0, GX~349+2, Cyg~X-2) and two accreting millisecond X-ray pulsars (SAX~J1808.4$-$3658, HETE~J1900.1$-$2455, which are also atolls).  The observations we analyze here were obtained with either \suz{} or \xmm{}, both of which have good effective area through the Fe K region (6.4 -- 6.97 keV).  In Table~\ref{tab:obs} we detail the observations presented here, as well as giving the distance to each source assumed throughout the paper.

\tabletypesize{\scriptsize}
\begin{deluxetable*}{lclccccccc}
\tablecolumns{10}
\tablewidth{0pc}
\tablecaption{Observation details}
\tablehead{Source & Class & Mission & Obs. ID & Obs. start date & Exp. time (ks) & Mode & Ref. & Distance (kpc) & Distance ref.}
\startdata
Serpens~X-1 & A & \suz & 401048010 & 24/10/2006 & 18 / 29 & 1/4 W, 1.0s & 1 & 8.4 & 11  \\
              &   & \xmm & 0084020401 & 22/03/2004 & 6 & PN: timing & 2  & &  \\
              &   & \xmm & 0084020501 & 24/03/2004 & 7 & PN: timing & 2  & & \\
              &   & \xmm & 0084020601 & 26/03/2004 & 7 & PN: timing & 2  &  & \\
4U~1636$-$53    & A & \xmm & 0303250201 & 29/08/2005 & 29 & PN: timing & 3 & $6.0\pm0.1$ & 12  \\
              &   & \xmm & 0500350301 & 28/09/2007 & 26 & PN: timing & 3 & &  \\
              &   & \xmm & 0500350401 & 27/02/2008 & 37 & PN: timing & 3 & &  \\
4U~1705$-$44  & A & \suz & 401046010 & 29/08/2006 & 14 / 14 & 1/4 W, 1.6s & 4 & $5.8\pm0.2$ & 12  \\
              &   & \suz & 401046020 & 18/09/2006 & 17 / 15 & 1/4 W, 2.0s & 4 & &  \\
              &   & \suz & 401046030 & 06/10/2006 & 18 / 17 & 1/4 W, 2.0s & 4 & & \\
              &   & \xmm & 0402300201 & 26/08/2006 & 34 & PN: timing & 5 & &  \\
4U~1820$-$30  &	A & \suz & 401047010 & 14/09/2006 & 10 / 30 & 1/4 W, 1.0s & 1 & $7.6\pm0.4$ & 13  \\
GX~17+2       & Z & \suz & 402050010 & 19/09/2007 & 5 / 15 & 1/4 W, 0.5s & 5 & $9.8\pm0.4$ & 12   \\
              &   & \suz & 402050020 & 27/09/2007 & 6 / 18 & 1/4 W, 0.5s & 5 & &   \\
GX~340+0      & Z & \xmm & 0505950101 & 02/09/2007 & 33 & PN: timing & 6 & $11\pm3$ & 14  \\
GX~349+2      & Z & \suz & 400003010 & 14/03/2006 & 8 / 20 & 1/8 W, 0.3s & 1 & 5 & 11  \\
              &   & \suz & 400003020 & 19/03/2006 & 8 / 24 & 1/8 W, 0.3s & 1 & &   \\
              &   & \xmm & 0506110101 & 19/03/2008 & 10 & PN: timing & 7  & &  \\
Cyg~X-2     & Z & \suz & 401049010 & 16/05/2006 & $<0.5$ & 1/4 W, 1.0s & 8 & $11\pm2$ & 12   \\
              & Z & \suz & 403063010 & 01/07/2008 & 20 / 82 & 1/4 W, 0.5s & 5 & &  \\
SAX~J1808.4$-$3658  & AMXP & \suz & 903003010 & 02/10/2008 & 21 / 31 & 1/4 W, 1.0s & 9 & $3.5\pm0.1$ & 15  \\
                    &      & \xmm & 0560180601 & 30/09/2008 & 43 & PN: timing & 9,10 & &  \\ 
HETE~J1900.1$-$2455 & AMXP & \suz & 402016010 & 16/10/2007 & 42 / 37 & 1/4 W, 2.0s & 5 & $3.6\pm0.5$ & 12   
\enddata
\tablecomments{A: atoll, Z: Z source, AMXP: accreting millisecond X-ray pulsar.  The \suz{} exposure times given are (1) the good time for the individual XIS detectors, (2) the good time for the HXD/PIN.  For \suz{} modes 'W' stands for window-mode (where a sub-array is used). The number following that is the frame time.  The nominal frame time for 1/4 W is 2.0s, and for 1/8 W is 1.0s, shorter frame times encur deadtime. Where no uncertainty in distance is given, we assume a 25\% uncertainty.}
\tablerefs{
(1) \citealt{cackett08},
(2) \citealt{bhattacharyya07},
(3) \citealt{pandel08},
(4) \citealt{reis09_1705},
(5) This work,
(6) \citealt{dai09},
(7) \citealt{iaria09},
(8) \citealt{shaposhnikov09},
(9) \citealt{cackett_j1808_09},
(10) \citealt{papitto09}, 
(11) \citealt{christian97},
(12) \citealt{galloway08},
(13) \citealt{kuulkers03},
(14) \citealt{fender00},
(15) \citealt{galloway06}}
\label{tab:obs}
\end{deluxetable*}

\section{Data Reduction}
\label{reduction}

Details of our data reduction for the \suz{} and \xmm{} observations follows.  We have attempted to follow a standard, common analysis procedure for all sources as much as possible.  Where we have had to deviate from this, specific details for those observations are also given. 

\subsection{\suz{} data reduction}

The data reduction was performed using HEASOFT v6.6.2, and the latest calibration files (as of June 2009).  \suz{} consists of both a set of soft X-ray CCD detectors (XIS) as well as a separate hard X-ray detector (HXD).  Details of the data reduction for the XIS and HXD are given below. For all sources we adopted the following standard data reduction method, except where explicitly stated.  Since our original analysis of Ser~X-1, GX~349+2 and 4U~1820$-$30 \citep{cackett08}, there has been significant advances in the analysis tools for \suz{}, for instance one can now generate response files for any source region (allowing for pile-up correction).

\subsubsection{XIS data reduction}
Firstly, we reprocessed the unfiltered event files using \verb!xispi! which updates the PI values for the latest calibration.  From there, we then created a cleaned, filtered event list using the standard screening criteria provided by \suz.  In addition to the standard screening we were also careful to filter out any periods where the telemetry was saturated.  This only occurred in a very small number of cases, which are noted below.  There are event lists for each data mode used on each detector.  We treated all of them separately until after spectra and responses were generated. 

There are a couple of exceptions to this procedure. When a source is bright, the script to remove flickering pixels (\verb!cleansis!) sometimes produces spurious results, rejecting all the brightest pixels in the image.  This was the case for a portion of both the second Cyg~X-2 observation and the Ser~X-1 observation.  To overcome this, one must run \verb!cleansis! without any iterations (as recommended in the \verb!cleansis! user guide).  We adopted this procedure for the few observations where this occurs.

From the cleaned, filtered event files we extracted spectra.  Source extraction regions were chosen to be a rectangular box of  $250 \times 400$ px for 1/4 window mode observations and $125 \times 400$ px for 1/8 window mode observations.  For the brightest sources where pile-up is present, we excluded a circular region centered on the source, the radius of which depended on severity of pile-up (radii are given below).

After extraction of spectra, we generated rmf and arf files using \verb!xisrmfen! and \verb!xissimarfgen! (using 200,000 simulated photons) for each spectrum.  We combined all the spectra and responses for the separate data modes (e.g., $2\times2$, $3\times3$) used for each detector using the \verb!addascaspec! tool.  Additionally, we then combined the spectra from all available front-illuminated detectors (XIS 0, 2 and 3).  The spectra were then rebinned by a factor of 4 to more closely match the HWHM of the spectral resolution.\\

We now give observation specific details for each source:\\

{\it Ser~X-1}: a central circular region of radius 30 px was excluded from the extraction region.  There is also a small amount of telemetry saturation.  We exclude times when this occurs, which corresponds to removing 3\% of XIS 0, 2 and 3 events, and 0.1\% of XIS 1 events.

{\it 4U~1705$-$44}: in the first and third \suz{} observations no pile-up correction was needed.  However, in the second observation, when the source was significantly brighter, a central circular region of radius 20 px was excluded.  In the first observation 2 type-I X-ray bursts were seen in the lightcurve, which we filtered out.  No bursts were detected in the other two observations.

{\it 4U~1820$-$30}: a central circular region of radius 40 px was excluded from the extraction region.  There is also some telemetry saturation, and we exclude times when this occurs.  This removes approximately 16\% of events on all XIS detectors.

{\it GX~17+2}: a central circular region of radius 30 px was excluded from the extraction region for both observations of this source.  Only XIS 0 and XIS 3 data are analyzed -- XIS 2 was not in operation, and XIS 1 was not operated in window mode.  We detected two X-ray bursts (one during each observation), which were excluded.

{\it GX~349+2}: a central circular region of radius 30 px was excluded from the extraction region for both observations of this source.  The XIS 0, 2 and 3 detectors were operated in 1/8 window mode, with a 0.3s integration time.  The XIS 1 detector, however, was operated in 1/4 window mode, but also with a 0.3s integration time, and thus has a smaller livetime fraction (15\% compared to 30\%).

{\it Cyg~X-2}: the first observation of this source has previously been published by \citet{shaposhnikov09}.  These authors noted that a large fraction of this observation was affected by telemetry saturation and thus only used data from times when high telemetry rates were utilized.  On re-analysis, we found that even during high telemetry rates the data still suffered telemetry saturation.  Strictly filtering the data using time where there was no telemetry saturation leads to less than 500s of good time for each detector.  Due to the resulting low count statistics, we do not analyze this observation further.

During the second observation, Cyg~X-2 happened to be observed in a particularly bright phase, and we therefore excluded a central circular region of radius 75 px (see later discussion of pile-up in Section \ref{sec:pileup}).  We only extract spectra from XIS 1 and XIS 3. XIS 0 was operated in `PSUM' (timing) mode.  This mode is not yet calibrated, thus the spectra cannot be used for analysis.  There was significant telemetry saturation during this observation, and we remove all times when this occurs.  This leads to the exclusion of 36\% of events from XIS 1 and 9\% from XIS 3.

{\it SAX~J1808.4$-$3658}: we use the spectra from \citet{cackett_j1808_09}, and do not reprocess the data here.

{\it HETE~J1900.1$-$2455}: we only use the sections of the data that were operated in 1/4 window mode.  Given the low source count rate, no pile-up correction was required.

\subsubsection{HXD/PIN data reduction}

In every case we downloaded the latest `tuned' background file from the \suz{} website.  The PIN spectrum was extracted from the cleaned event file using good time intervals common to both the PIN event file and the `tuned' background file.  The resulting source spectrum was then deadtime corrected.  The non-X-ray background spectrum was extracted from the `tuned' background file, using the same common good time intervals.  In addition to the non-X-ray background, the cosmic X-ray background also contributes a small ($\sim 5\%$) amount to the total background.  We fake a cosmic X-ray background spectrum following the standard \suz{} procedure, and add this to the non-X-ray background spectrum to create the total background spectrum.  The response of the detector has changed over time, thus, we make sure that we used the correct response associated with the epoch of the observation.  In the few cases where type-I X-ray bursts were detected in the XIS lightcurves, we also filtered out the same times from the HXD/PIN data.

\subsection{\xmm{} data reduction}

We processed the observation data files (ODFs) for each observation using the \xmm{} Science Analysis Software (v8).  For the analysis here we only use timing mode data from the PN camera as it is the appropriate observing mode for the sources of interest.  Calibrated event lists were created from the ODFs using the PN processing tool \verb!epproc!.  Exceptions to this procedure are the observations of SAX~J1808.4$-$3658 and GX~340+0.  We obtained the calibrated event lists for these observations directly from the \xmm{} SOC.

Before extracting spectra, we checked for periods of high background by extracting a lightcurve from an off-source strip with energy $> 10$ keV.  We note where we found high background levels below.  Using only times of low background, we extracted spectra with the following criteria: quality flag = 0, pattern $\leq 4$ (singles and doubles) and in the energy range 0.3 -- 12 keV.  The source extraction regions used were 24 pixels wide (in RAWX), covered all RAWY values and were centered on the brightest RAWX column.  In a few cases, where the source was particularly bright, we had to exclude the brightest two RAWX columns (details noted below).  The corresponding rmf and arf were created using the \verb!rmfgen! and \verb!arfgen! tools following the SAS procedures.

Observation specific details follow:

{\it Ser~X-1}: during the first observation, there is only a slightly high background at the beginning, which we choose to include. There is no background flaring in the other two observations.

{\it 4U~1636$-$53}: there were three X-ray bursts during the first observation, one during the second observation, and one during the third observation, all of which were excluded. Only during the second observation was there some background flaring.  Removing this reduced the exposure time by 3.5 ks. 

{\it 4U~1705$-$44}:  one type-I X-ray burst occurred during this observation, and was excluded.  We find no background flaring. Note that this is {\it not} the dataset presented in \citet{disalvo09}, which is still proprietary at the time of writing.  Other archival \xmm{} observations of this source were observed in imaging mode, and so are not analyzed here. 

{\it GX~340+0}: some low-level background flaring, which reduces the exposure time by 7 ks.  The source is particularly bright during this observation \citep[see lightcurve in][]{dai09}, we therefore remove the two brightest columns from the extract region. Given that \cite{dai09} demonstrate that the Fe K line properties do not vary significantly with source state, we chose not to split the observation based on source state, but instead prefer to analyze the entire time averaged spectrum.

{\it GX~349+2}:  the source count rate is high during this observation \citep{iaria09}.  Following \citet{iaria09}, we therefore exclude the two brightest columns from the extraction region.  There is no significant background flaring.  Another archival observation of this source is also available, however, it was observed in imaging mode, and thus is not analyzed here.

{\it SAX~J1808.4$-$3658}: we use the spectra from \citet{cackett_j1808_09} and do not reprocess the data here.

\section{Spectral Fitting}
\label{specfit}

\subsection{Phenomenological models}

Our first step is to fit the spectra with phenomenological models \citep[throughout the paper we use XSPEC v12;][]{arnaud96}.  There has been debate for many years as to the most appropriate continuum model to fit to neutron star LMXBs \citep[e.g][]{white88,mitsuda89}, as it is often the case that several different models fit the data equally well \citep[e.g][]{barret01}.  A recent investigation by \citet{lin07} studied two atoll sources over a wide range in luminosity.  They investigated the luminosity dependence of measured temperatures for different continuum models, looking to see for which model choice the measured temperatures followed $L \propto T^4$.  Based on this, they suggested a prescription for spectral fitting, using an absorbed disk blackbody \citep[xspec model: diskbb][]{arnaud96} plus single temperature blackbody plus a (broken) power-law for the soft and intermediate states, and a blackbody plus broken power-law for the hardest states.  We have also found this model to fit the spectra of neutron star LMXBs well \citep{cackett08,cackett_j1808_09,cackett_chazss_09}.  Here, we adopt this same model (though use a power-law, rather than a broken power-law), testing that the addition of each component statistically improves the fit.

We interpret the hot blackbody component as the boundary layer between the accretion disk and neutron star surface \citep[e.g.][]{inogamov99,popham01}.  One can also successfully model this component with a Comptonized model, such as \verb!comptt!.  When doing this, however, one gets high optical depths -- the Comptonized component is close to a Wien spectrum. Rather than fit the more complicated Comptonization model, we chose the simpler blackbody model. 

The millisecond X-ray pulsar, HETE J1900.1$-$2455 has a much harder spectrum than the other sources we study here.  In fact, we find that its broadband continuum spectrum can be well fit be a single power-law only.

We found that several of the sources have an apparent emission feature at around 1 keV, as has previously been observed in a number of neutron star LMXBs, with a number of different missions \citep[e.g.][]{vrtilek88,kuulkers97,schulz99}.  It has suggested that this feature is a blend of a number of emission lines from Fe and O \citep{vrtilek88}.  Here, if the feature is present, we fit with a single Gaussian emission line.

The presence of a broad Fe K emission line in eight of these sources has already been established elsewhere (see references earlier).  Of the other two sources (GX~17+2 and HETE~J1900.1$-$2455) a  Gaussian Fe K emission has previously been reported in GX~17+2 \citep{disalvo00,farinelli05,cackett_chazss_09} and the fact that a broad red-wing has not been previously observed is likely due to sensitivity \citep{cackett_chazss_09}. Here we show that the profile is asymmetric.  In HETE~J1900.1$-$2455, we detect a broad Fe K emission line for the first time, though the line can be equally well fit by a broad Gaussian or a relativistic line due to the low signal-to-noise ratio.  Therefore, all 10 objects studied here are seen to have broad Fe K lines.

To fit these emission lines, we use the relativistic line model for a Schwarzschild metric \citep[diskline;][]{fabian89}.  The Schwarzschild metric should be a good approximation for neutron stars -- for neutron stars with reasonable equations of state and spin frequencies less than 600 Hz, the dimensionless angular momentum parameter is expected to be less than 0.3, thus any deviations from the Schwarzschild metric are minor \citep[see][for further discussion of this]{miller98_rot}.  We note that the diskline profile does not account for light bending, and so does not fully account for all relativistic effects.  However, it still remains the most appropriate available model here.

We include this model in addition to the continuum model discussed above.  The emission line parameters are the line energy, disk emissivity index, inner disk radius, outer disk radius and inclination.  We constrain the line energy to be within 6.4 to 6.97 keV, the allowed range for Fe K$\alpha$ emission.  The disk emissivity index and inner disk radius are left as free parameters in the fits.  In most cases, the inclination is left as a free parameter (we discuss the exceptions to this later).  Finally, we fix the outer disk radius at 1000 $GM/c^2$.  Typical disk emissivities mean that the Fe K emission is quite centrally concentrated, and hence the outer disk radius is usually not well constrained, and has little affect on the measured inner disk radius.

Several of the sources considered here have multiple observations by the same, or different telescopes.  The inclination of the source should not change over time, thus, we should expect to recover the same inclination from the Fe K line modeling for each observation.  In order to investigate the extent to which this is true, we chose to let the inclination parameter vary from observation to observation.

For the \suz{} spectra, we fit the XIS spectra over the energy range 1.0 -- 10 keV, ignoring 1.5 -- 2.5 keV where large residuals are always seen (regardless of count rate), presumably of an instrumental nature.  The PIN spectra are fit from 12 keV upwards, with the upper energy depending on where the background dominates over the source spectrum.  The combined front-illuminated spectrum, XIS 1 spectrum and HXD/PIN spectra are all fit simultaneously, with a floating constant used to account for any absolute flux calibration offset between them. When fitting the \xmm{} PN spectra, we fit over the 0.8 -- 10 keV range.  Source specific details, including exceptions to this standard procedure follow.

{\it Ser~X-1}: \suz{}/PIN spectrum fit from 12 -- 25 keV.  In the \xmm{} spectra there are significant residuals around 1 keV that are well fit by a Gaussian emission line.

{\it 4U~1636$-$53}: The Fe line profile observed in this source \citep{pandel08}, is unlike what is seen in any of the other sources.  It has a blue wing that extends to high energies, suggestive of a high inclination. Our spectral fitting does give a best fit with a high inclination (we find that all cases peg at 90$^\circ$) and \citet{pandel08} also reported high inclinations (greater than 64$^\circ$ in all cases).  However, no eclipses have been detected in this source and  optical measurements are not especially constraining \cite[$36^\circ$ -- $74^\circ$,][]{casares06}.

{\it 4U~1705$-$44}: \suz{}/PIN spectrum fit from 12 -- 30 keV.  The first \suz{} observation has quite low sensitivity through the Fe K band \citep[see][]{reis09_1705}.  We therefore fix several of the line parameters. 

When fitting the \xmm{} data of 4U~1705$-$44, we fit in the range 1.2 -- 10 keV, as a relatively high $N_{\rm H}$ and low source flux lead to low statistics below 1.2 keV.  We found that both the diskbb+bbody or bbody+power-law model fit the continuum spectra well, with the diskbb+bbody model having a marginally better $\chi^2$ value.  The spectrum is of fairly low quality through the Fe K region, and thus a simple Gaussian models the Fe K line well.  However, to allow comparison with other observations we still fit the line with the diskline model. When including the diskline model both choices of continuum model give consistent Fe K parameters, with almost identical $\chi^2$ values: 1762.4 vs. 1762.5 (dof = 1752 in both cases) for diskbb+bbody and bbody+power-law continuum choices respectively.

An unphysically high inclination is determined from the \xmm{} observation.  This source is not known to be eclipsing or dipping. There is a different inclination determined in the \suz{} observations \citep{reis09_1705} than from \citet{disalvo09}.  We revisit this issue later when discussing systematic uncertainties in Fe K modeling.

{\it 4U~1820$-$30}: \suz{}/PIN spectrum fit from 12 -- 25 keV.  With the improved calibration, the combined front-illuminated spectrum has a significantly higher signal-to-noise ratio than the spectra presented by \citet{cackett08}.  With this improved spectrum, the detected emission line is found to be weaker than before, but still a significant detection: with the emission line the best fit has $\chi^2 = 1269$ for 1108 degrees of freedom, whereas with no emission line the best fit has $\chi^2 = 1362$ for 1113 degrees of freedom.

{\it GX~17+2}: \suz{}/PIN spectrum fit from 12 -- 40 keV. 

{\it GX~340+0}: This source is hampered by a very large column density ($N_{\rm H} \sim 10^{23}$ cm$^{-2}$).  This means that even the neutral Fe K edge at $\sim$7 keV is significant.  Unfortunately, a large neutral Fe K edge can be a big problem when fitting the Fe K emission line.  If there is any small change in ISM Fe abundance towards the source, it will change the depth of the edge, and consequently the appearance of the emission line.  Additionally, the edge is likely to be more complicated  than the simple step-like function used in absorption models.  Detailed structure has been observed for Fe L, and several Ne and O edges with sensitive gratings spectra \citep[e.g.][]{juett04,juett06}.  Thus, with a large $N_{\rm H}$, structure in the edge might affect results.

\begin{figure}
\centering
\includegraphics[width=7.5cm]{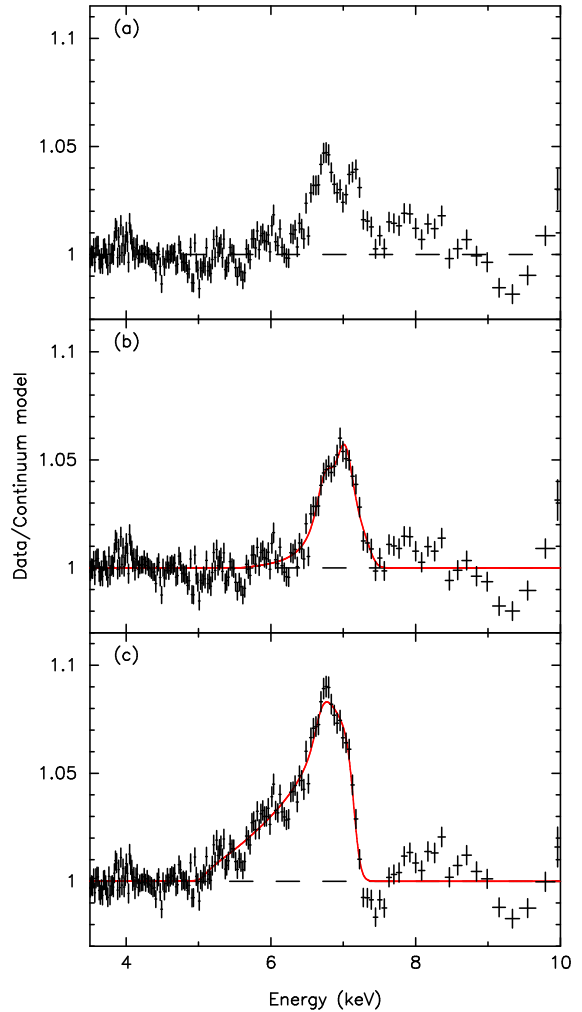}
\caption{Ratio of data to continuum model  for GX~340+0.  Each panel shows the same spectrum, fit with a different model (see Table~\ref{tab:gx340} for spectral fits). (a) The continuum is fit from 2.5-5 keV and 8-10 keV, ignoring the Fe K region, with an absorbed blackbody + disk blackbody model.  A clear double-peaked structure is apparent. (b) An absorption line at 6.97 keV is added to the model along with a relativistic Fe K emission line.  The ratio shown is that of the data to the continuum + absorption line model, thus, showing the emission line profile only.  The red, solid line shows the emission line model. (c) Here we follow \citet{dai09}, allowing the ISM Fe abundance to vary, changing the depth of the Fe K edge from ISM absorption.  The emission line model (red, solid line) is shown.}
\label{fig:gx340}
\end{figure}

When fitting these data, \citet{dai09} chose to allow the ISM Fe abundance to vary, effectively changing the size of the Fe K edge.  This produces a good fit, and the resulting Fe K emission line profile looks good.  We, similarly, achieve good fits by varying the Fe abundance.  Nevertheless, there is not a unique model that fits the data well (see the line profiles in Fig.~\ref{fig:gx340}). Fig.~\ref{fig:gx340}a shows the resulting line profile when the continuum is fit from 2.5 -- 5 keV and 8 -- 10 keV, ignoring the Fe K region (and assuming standard abundances).  Evidently, the resulting residuals do not look like a typical Fe K emission line profile, with a clear dip at around 7 keV.  As mentioned above, \citet{dai09} successfully correct for this by assuming that the ISM Fe abundance towards this source is lower (see Fig.~\ref{fig:gx340}c).  With a lower Fe abundance, the absorption edge is weaker, thus the line profile no longer shows a dip.

Here, we alternatively suggest that this can also be interpreted as an \ion{Fe}{26} absorption line at 6.97 keV superposed on a broad Fe K emission line (see Fig.~\ref{fig:gx340}b, and Table~\ref{tab:gx340}).  There are precedents for such a scenario.  For instance, \citet{boirin05} find \ion{Fe}{25} and \ion{Fe}{26} absorption lines superposed on a broad Fe K emission line in the dipping LMXB 4U~1323$-$62, and recently similar absorption lines superposed on a broad Fe K emission line were observed in the black hole binary GRO~J1655$-$40 \citep{diaztrigo07}.

As is apparent in Fig.~\ref{fig:gx340}, the two models lead to differing emission line profiles.  While in the model varying Fe abundance the line profile looks asymmetric, it is not as obviously asymmetric when including an absorption line.  Fitting the emission line profiles with the \verb|diskline| model we also achieve different measurements of $R_{\rm in}$:  $9.3\pm0.2~GM/c^2$ \citep[consistent with the analysis of][]{dai09} compared to $22\pm4~GM/c^2$ for the alternative model with an absorption line included.  Due to the ambiguous nature of the line profile in this source we do not consider this spectrum further here, though parameters from the above spectral fits can be found in the Table~\ref{tab:gx340}.

\tabletypesize{\scriptsize}
\begin{deluxetable}{lcc}
\tablecolumns{3}
\tablewidth{0pc}
\tablecaption{Spectral fitting parameters for GX~340+0}
\tablehead{Parameter & Variable Fe abundance & Fe XXVI Absorption line}
\startdata
$N_H (10^{22}$ cm$^{-2}$) & $10.9\pm0.1$ & $10.0\pm0.1$ \\
Fe abundance (rel. to solar) & $0.25_{-0.07}^{+0.03}$ & 1.0 (fixed) \\
Disk $kT$ (keV) & $1.64\pm0.04$ & $1.21\pm0.02$ \\
Disk normalization & $127\pm7$ & $325\pm9$ \\
Blackbody $kT$ (keV) & $2.44\pm0.09$ & $1.93\pm0.01$ \\
Bbody normalization ($10^{-2}$) & $9.2\pm0.5$ & $15.2\pm0.1$ \\
Gaussian line energy (keV) & -- & 6.97 (frozen) \\
Gaussian $\sigma$ (keV) & -- & 0.05 (frozen) \\
Gaussian normalization & -- & $(-1.1\pm0.1)\times10^{-3}$ \\
Diskline line energy (keV) & $6.97_{-0.01}$ & $6.89\pm0.01$ \\
Emissivity, q & $3.7\pm0.1$ & $1.85\pm0.14$ \\
$R_{\rm in} (GM/c^2)$ & $9.3\pm0.2$ & $22\pm4$ \\
$i (^\circ)$ & $22\pm1$ & $37\pm4$ \\
Diskline normalization $(10^{-3})$ & $13.1\pm0.2$ & $5.5\pm0.2$ \\
EW (eV) & 80 & 38 \\
$\chi^2/dof$ & 1827/1548 & 1925/1548
\enddata
\tablecomments{The outer disk radius was fixed at 1000 $GM/c^2$ in both cases. Normalization of the disk blackbody is in $R_{\rm inner}^2\cos{i} /(D10)^2$ where $R_{\rm inner}$ is the inner disk radius in km, $D10$ is the distance to the source in units of 10 kpc, and $i$ is the disk inclination. Normalization of the blackbody component is $L39/(D10)^2$, where $L39$ is the luminosity in units of $10^{39}$ erg s$^{-1}$. The normalization of both the Gaussian and the diskline is the line flux in photons cm$^{-2}$. s$^{-1}$.}
\label{tab:gx340}
\end{deluxetable}

{\it GX~349+2}: We fit the two \suz{} spectra separately, even though the observation times are close together, in order to investigate any variability.  The \suz{}/PIN spectrum was fit from 12 -- 25 keV. In the \xmm{} spectrum, there are large residuals at $\sim$ 1 keV, that are well fit by a Gaussian emission line \citep{iaria09}.

{\it Cyg~X-2}: XIS spectra fitted from 0.8 -- 10 keV, ignoring 1.5 -- 2.5 keV.  The \suz{}/PIN spectrum was fit from 12 -- 25 keV.  We find large residuals at $\sim$ 1 keV (in all XIS spectra), that are well fit by a single Gaussian emission line.  This spectral feature has previously been observed in this source \citep[e.g.][]{vrtilek88,piraino02,shaposhnikov09}, and may be a complex of blended emission lines \citep{vrtilek88,schulz09}.

{\it SAX~J1808.4$-$3658}: we do not re-fit the \suz{} and \xmm{} spectra of this source here, and refer the reader to \citet{cackett_j1808_09} regarding specifics.

{\it HETE~J1900.1$-$2455}: The \suz{}/PIN spectrum was fit 12 -- 30 keV.  This spectrum is quite noisy, though an Fe K emission line is apparent.  A broad ($\sigma=0.8$ keV) Gaussian fits the line well ($\chi^2 = 1148$, d.o.f. = 1228), and a relativistic disk line is not statistically required.  We do, however, fit the diskline to allow comparison with other data, and it gives a similarly good fit ($\chi^2 = 1146$, d.o.f. = 1226).  Also note that the continuum model for this source only requires a power-law component and no disk blackbody, or blackbody are required. 

\subsubsection{Investigating the effects of pile-up}\label{sec:pileup}

\begin{deluxetable*}{lcccc}
\tablecolumns{5}
\tablewidth{0pc}
\tablecaption{Investigating pile-up in Cyg~X-2: spectral parameters for different extraction regions}
\tablehead{Parameter & \multicolumn{4}{c}{Excluded circle radius}\\
 & 0 px & 30 px & 60 px & 90 px}
\startdata
$N_H (10^{22}$ cm$^{-2}$) & $0.147\pm0.003$ & $0.136\pm0.004$ & $0.128\pm0.004$ & $0.130\pm0.005$ \\
Disk $kT$ (keV) & $1.595\pm0.006$ & $1.555\pm0.002$  & $1.52\pm0.02$ & $1.45\pm0.04$  \\
Disk normalization & $124\pm2$ & $151\pm1$ & $165\pm3$ & $190\pm20$ \\
Blackbody $kT$ (keV) & $9\pm1$ & $4.1\pm0.1$ & $2.9\pm0.2$ & $2.2\pm0.2$ \\
Bbody normalization $(10^{-2})$ & $41\pm2$ & $7.0\pm0.1$ & $4.2\pm0.1$ & $4.4\pm0.6$ \\
Line Energy (keV) & $6.97_{-0.02}$ & $6.97_{-0.02}$ & $6.97_{-0.02}$ & $6.97_{-0.02}$ \\
Emissivity, q & $5.8\pm0.6$ & $5.7\pm0.5$ & $6.3\pm0.8$ & $6.5^{+1.0}_{-1.5}$ \\
$R_{\rm in}$ ($GM/c^2$) & $8.0\pm0.2$ & $7.8\pm0.2$ & $7.9\pm0.3$ & $8.0\pm0.5$ \\
$i$ ($^\circ$) & $24.0\pm0.5$ & $24.3\pm0.6$ & $23.9\pm0.6$ & $23\pm1$ \\
Line normalization $(10^{-3})$ & $7.2\pm0.4$ & $8.4\pm0.4$ & $8.7\pm0.6$ & $8.4\pm1.0$ \\
EW (eV) & 69 & 76 & 79 & 74 
\enddata
\tablecomments{Spectral parameters from fitting the XIS 3 data. While the continuum is variable due to pile-up effects, the Fe K line parameters are remain consistent. Normalization of the disk blackbody is in $R_{\rm inner}^2\cos{i} /(D10)^2$ where $R_{\rm inner}$ is the inner disk radius in km, $D10$ is the distance to the source in units of 10 kpc, and $i$ is the disk inclination. Normalization of the blackbody component is $L39/(D10)^2$, where $L39$ is the luminosity in units of $10^{39}$ erg s$^{-1}$. The normalization of the diskline is the line flux in photons cm$^{-2}$. s$^{-1}$.}
\label{tab:cygx2pileup}
\end{deluxetable*}

\begin{figure*}
\centering
\includegraphics[width=7.5cm]{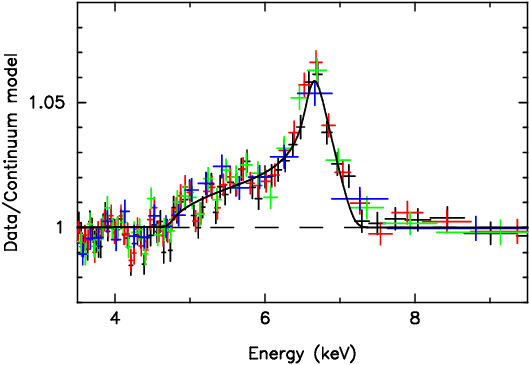}
\hspace{0.5cm}
\includegraphics[width=7.5cm]{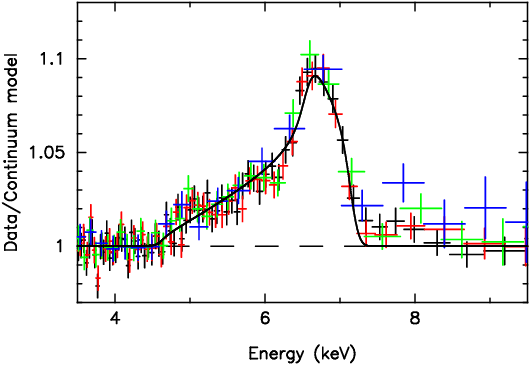}
\caption{{\it Left:} Fe line profile for Cyg~X-2, excluding a central circular region of varying radius. Black: exclusion circle radius = 0 px, red: radius = 30 px, green: radius = 60 px, blue: radius = 90 px.  The Fe K line profile remains consistent regardless of the extraction region used. {\it Right} Fe line profile for Ser~X-1 from the {\it Suzaku} observation.  Only data from XIS 3 is shown.  Colors denote the same size extraction regions as for Cyg~X-2.}
\label{fig:cygx2pileup}
\end{figure*}

Of all the sources observed, Cyg~X-2 had the highest count rate observed by \suz, leading to severe pile-up. The continuum shape changes due to pile-up, becoming artificially harder \citep[e.g.][]{davis01}.  In order to investigate the robustness of the Fe K line profile in Cyg~X-2, we extracted spectra from multiple different regions.  In all cases, we used a box $250\times400$ px, but had a circular exclusion region of varying radius centered on the source.  The radius of this circle ranged from 0 px to 90 px.  We fitted the spectra from 0.8 -- 10 keV, ignoring the 1.5 -- 2.5 keV energy range, as usual.  The spectra were allowed to have completely different continuum models, fit using an absorbed disk blackbody plus a single temperature blackbody.  Additionally, we had to include a narrow Gaussian at 1 keV.   Here, we do not fit the PIN data, thus no power-law component is required.  We do not fit the PIN data as pile-up significantly effects the shape of the XIS spectrum, hence the broadband spectrum will not fit well unless the XIS spectra are corrected for pile-up.

We found that even though each spectrum had different continuum shapes (due to pile-up effects), the Fe K line profile was remarkably robust.  This is clearly demonstrated in Tab.~\ref{tab:cygx2pileup} and Fig.~\ref{fig:cygx2pileup}.  In Tab.~\ref{tab:cygx2pileup} we give the continuum and Fe K line parameters from fitting the spectra with different size exclusion regions.  Fig.~\ref{fig:cygx2pileup} shows the resulting line profiles for each spectrum.  The Fe K line is both quantitatively and qualitatively consistent, regardless of the pile-up correction.  Thus, while with pile-up the continuum may vary significantly from the `true' (not piled-up) continuum shape (in the most extreme cases it is obvious that the continuum model parameters are not realistic), it appears that the Fe K line shape can be robustly recovered as long as the continuum can be fit well.   To further demonstrate this, we also show the Fe K profile from the {\it Suzaku} observation of Ser~X-1 (Fig. \ref{fig:cygx2pileup}).  We again see that the line profile remains consistent as the radius of the circular exclusion region is varied from 0 px to 90 px.

\subsubsection{Results from phenomenological fitting}

\begin{figure*}
\centering
\includegraphics[angle=-90,width=\textwidth]{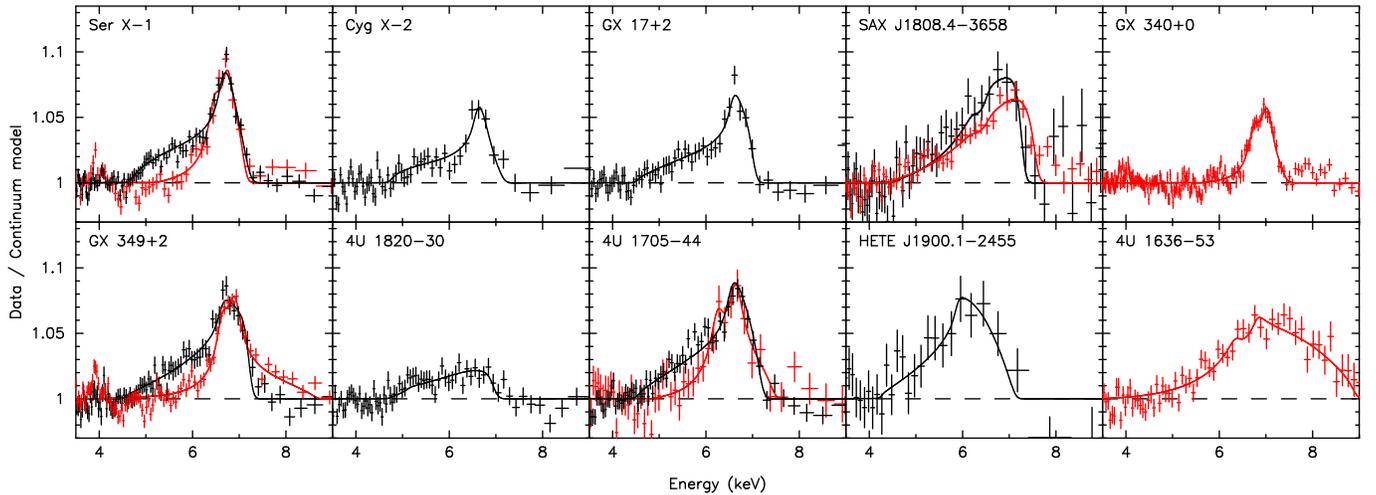}
\caption{A summary of the Fe K emission lines in neutron star LMXBs.  Plotted is the ratio of the data to the continuum model.  The solid line shows the best-fitting diskline model.  Data from \suz{} is shown in black (combined front-illuminated detectors), data from \xmm{} is shown in red (PN camera).}
\label{fig:diskline}
\end{figure*}

The spectral parameters from this phenomenological modeling are given in Tables~\ref{tab:phenom_cont} and \ref{tab:phenom_line} (the tables have been split into continuum and line parameters).  Table~\ref{tab:fL} gives the 0.5 -- 25 keV unabsorbed fluxes and luminosities.   A summary of all the Fe K lines observed in neutron star LMXBs can be seen in Fig.~\ref{fig:diskline}.

\tabletypesize{\tiny}
\begin{deluxetable*}{lclcccccccccc}
\tablecolumns{13}
\tablewidth{0pc}
\tablecaption{Phenomenological model parameters: continuum}
\setlength{\tabcolsep}{0.04in}
\tablehead{Source & Mission & Obs. ID & $N_H$ & \multicolumn{2}{c}{Disk Blackbody} & \multicolumn{2}{c}{Blackbody} & \multicolumn{2}{c}{Power-law}  & \multicolumn{3}{c}{Gaussian} \\
 & & & $10^{22}$ cm$^{-2}$ & $kT_{\rm in}$ (keV) & Norm. & $kT$ (keV) & Norm. $(10^{-2})$ & $\Gamma$ & Norm. & Line energy (keV) & $\sigma$ (keV) & Norm. $(10^{-2})$}
\startdata
Serpens~X-1  & \suz & 401048010 &  $0.67\pm0.04$  & $1.26\pm0.01$ & $97\pm5$  & $2.27\pm0.02$ & $5.23\pm0.06$ & $3.4\pm0.3$ & $0.9\pm0.2$ & -- & -- & --  \\
               & \xmm & 0084020401 & $0.35\pm0.01$  & $0.95\pm0.01$ & $205\pm1$ & $1.76\pm0.01$ & $4.46\pm0.01$ & -- &	-- & $1.10\pm0.01$ & $0.11\pm0.01$ & $1.3\pm0.01$  \\
               & \xmm & 0084020501 & $0.35\pm0.01$  & $0.99\pm0.01$ & $175\pm3$ & $1.81\pm0.01$ & $3.13\pm0.02$ & -- &	-- & $1.10\pm0.01$ & $0.14\pm0.01$ & $1.4\pm0.02$ \\
               & \xmm & 0084020601 & $0.35\pm0.01$  & $0.95\pm0.01$ & $203\pm2$ & $1.76\pm0.01$ & $3.80\pm0.01$ & -- &	-- & $1.10\pm0.01$ & $0.13\pm0.01$ & $1.2\pm0.01$ \\
4U~1636$-$53     & \xmm & 0303250201 & $0.36\pm0.01$ & $0.21\pm0.01$ & $(1.0\pm0.2)\times10^4$ & $1.97\pm0.03$   & $0.15\pm0.01$  & $1.9\pm0.1$ & $0.22\pm0.01$ & -- & -- & -- \\
               & \xmm & 0500350301 & $0.51\pm0.01$ & $0.89\pm0.01$ & $101\pm4$               & $2.00\pm0.01$   & $1.19\pm0.01$	& $3.8\pm0.1$ & $0.54\pm0.01$ & -- & -- & --  \\
               & \xmm & 0500350401 & $0.50\pm0.01$ & $0.96\pm0.01$ & $103\pm4$               & $2.03\pm0.01$   & $1.57\pm0.01$	& $3.8\pm0.1$ & $0.58\pm0.01$ & -- & -- & --  \\
4U~1705$-$44   & \suz & 401046010 & 2.0 & $0.09\pm0.01$ & $(2.3_{-1.6}^{+7.2})\times10^{7}$ & $1.05\pm0.05$ & $0.063\pm0.005$  & $1.65\pm0.01$ & $0.124\pm0.002$ & -- & -- & --  \\
               & \suz & 401046020 &  $1.95\pm0.05$ & $1.16\pm0.02$ & $56\pm2$    & $2.12\pm0.03$ & $2.40\pm0.02$ & $2.85\pm0.08$ & $0.64\pm0.06$   & -- & -- & --  \\
               & \suz & 401046030 &  $2.02\pm0.05$ & $0.82\pm0.02$ & $82\pm5$    & $1.95\pm0.03$ & $0.95\pm0.01$ & $3.00\pm0.06$ & $0.49\pm0.03$  & -- & -- & --   \\
               & \xmm & 0402300201 & $1.53\pm0.01$ & $1.44\pm0.03$ & $2.5\pm0.2$ & $3.2\pm0.3$   & $0.25\pm0.01$ & --	         & --              & -- & -- & --  \\
4U~1820$-$30   & \suz & 401047010  & $0.21\pm0.01$ & $1.18\pm0.01$ & $103\pm5$   & $2.44\pm0.01$  & $7.7\pm0.1$  & $2.3\pm0.1$ &	$0.37\pm0.03$  & -- & -- & --  \\
GX~17+2        & \suz & 402050010 & $2.2\pm0.1$    & $1.75\pm0.02$ & $86\pm5$    & $2.66\pm0.02$  & $6.1\pm0.2$  & $2.6\pm0.6$ & $0.4\pm0.4$  & -- & -- & -- \\
               & \suz & 402050020 & $2.14\pm0.05$  & $1.74\pm0.02$ & $92\pm4$    & $2.55\pm0.02$ &  $9.2\pm0.3$  & $2.4\pm0.3$ & $0.2\pm0.2$  & -- & -- & -- \\
GX~349+2       & \suz & 400003010 & $0.77\pm0.01$  & $1.53\pm0.02$ & $112\pm4$   & $2.24\pm0.02$ &  $12.5\pm0.2$ & $1.3\pm0.7$ & $0.01_{-0.01}^{+0.05}$  & -- & -- & -- \\
               & \suz & 400003020 & $0.85\pm0.02$ & $1.48\pm0.02$  & $105\pm4$   & $2.30\pm0.01$ &  $9.6\pm0.1$ & $2.3\pm0.1$ & $0.31\pm0.01$ & -- & -- & -- \\
               & \xmm & 0506110101 & $0.67\pm0.01$ & $1.17\pm0.01$ & $258\pm6$   & $1.85\pm0.01$ &  $16.7\pm0.3$ & -- & -- &  $1.08\pm0.01$ & $0.09\pm0.01$ & $3.9\pm0.4$  \\
Cyg~X-2      & \suz & 403063010 & $0.20\pm0.02$  & $1.50\pm0.01$ & $163\pm4$   & $2.21\pm0.02$ &  $3.6\pm0.1$  & $2.8\pm0.1$ & $60\pm20$ & $1.01\pm0.01$ & $0.11\pm0.01$ & $9.0\pm0.6$  \\
SAX~J1808.4$-$3658   & \suz & 903003010 & $0.046\pm0.04$ & $0.48\pm0.01$ & $5.9\pm0.3$ & $1.00\pm0.02$ & $0.127\pm0.001$ & $1.93\pm0.01$ & $0.20\pm0.01$ & -- & -- & -- \\
                     & \xmm & 0560180601 & $0.23\pm0.01$ & $0.23\pm0.01$ & $20400^{+13200}_{-5000}$ & $0.40\pm0.01$ & $0.125\pm0.004$ & $2.08\pm0.01$ & $0.33\pm0.01$ & -- & -- & --  \\ 
HETE~J1900.1$-$2455  & \suz & 402016010 &  $0.032\pm0.04$ & -- &   -- &   -- &   --  &  $2.14\pm0.01$ & $0.063\pm0.001$ & -- & -- & --   \\
\enddata
\tablecomments{Continuum spectral fitting parameters from phenomenological modeling. All uncertainties are 1$\sigma$. Where no uncertainty is given, that parameter was fixed.  The $\chi^2$ values for the fits are those in Tab.~\ref{tab:phenom_line}. Normalization of the disk blackbody is in $R_{\rm inner}^2\cos{i} /(D10)^2$ where $R_{\rm inner}$ is the inner disk radius in km, $D10$ is the distance to the source in units of 10 kpc, and $i$ is the disk inclination. Normalization of the blackbody component is $L39/(D10)^2$, where $L39$ is the luminosity in units of $10^{39}$ erg s$^{-1}$.  Normalization of the power-law component gives the photons keV$^{-1}$ cm$^{-2}$ s$^{-1}$ at 1 keV.  Finally, the normalization of the Gaussian is the total number of photons cm$^{-2}$ s$^{-1}$ in the line.}
\label{tab:phenom_cont}
\end{deluxetable*}

\tabletypesize{\tiny}
\begin{deluxetable*}{lclccccccc}
\tablecolumns{10}
\tablewidth{0pc}
\tablecaption{Phenomenological model parameters: Fe K line}
\tablehead{Source & Mission & Obs. ID & \multicolumn{6}{c}{Diskline} & $\chi^2$/dof \\
 & &  & Line Energy (keV) & Emissivity index & $R_{\rm in}$ ($GM/c^2$) & Inclination $(^\circ)$ & Norm. $(10^{-3})$ & EW (eV) &  }
\startdata
Serpens~X-1  & \suz & 401048010 & $6.97_{-0.02}$ & $4.8\pm0.3$ & $8.0\pm0.3$ & $24\pm1$ & $5.6\pm0.4$ & 98 & 1442/1105 \\
               & \xmm & 0084020401 & $6.66\pm0.02$ & $1.8\pm0.2$ & $25\pm8$ & $32\pm3$ & $1.8\pm0.1$ &	43 & 2125/1831 \\
               & \xmm & 0084020501 & $6.97_{-0.03}$ & $3.7\pm4$ & $14\pm1$ & $13\pm2$ & $1.6\pm0.1$ & 50 &1961/1831 \\
               & \xmm & 0084020601 & $6.66\pm0.02$ & $2.0\pm0.2$ & $26\pm8$ & $27\pm2$ & $1.4\pm0.1$ & 38 &2151/1831 \\
4U~1636$-$53     & \xmm & 0303250201 & $6.4^{+0.02}$ & $2.7\pm0.1$ & $6.1^{+0.4}$	& $>80$ & $1.2\pm0.9$	& 184 & 1939/1832 \\
               & \xmm & 0500350301 & $6.4^{+0.04}$ & $2.8\pm0.1$ & $7.7\pm0.6$	& $>77$	& $1.2\pm0.1$	& 106 & 2112/1832\\
               & \xmm & 0500350401 & $6.4^{+0.02}$ & $2.8\pm0.1$ & $6.0^{+0.2}$	& $>84$	& $2.5\pm0.1$	& 164 & 2406/1832\\
4U~1705$-$44   & \suz & 401046010 & $6.62\pm0.03$  & 3.0  & 10.0  &  24   & $<0.44$ & $<76$  & 953/1122  \\
               & \suz & 401046020 & $6.94\pm0.06$ & $4.8\pm0.6$ & $7.3\pm0.4$ & $24\pm4$ & $3.3\pm0.3$ & 120 & 1266/1120  \\
               & \suz & 401046030 & $6.9\pm0.1$ &  $3.7\pm0.3$ &  $6^{+0.2}$  &  $15\pm3$ & $1.3\pm0.1$ & 94 & 1152/1120\\
               & \xmm & 0402300201 & $6.43\pm0.03$ & $2.1\pm0.1$ & $6^{+1}$ & $>74$ & $0.34\pm0.05$ & 151 & 1760/1752  \\
4U~1820$-$30   & \suz & 401047010 & $6.97-_{0.03}$ & $3.3\pm0.1$ & $6.0^{+0.1}$ & $<8$ & $2.6\pm0.3$ & 36 & 1269/1108\\
GX~17+2        & \suz & 402050010 & $6.8\pm0.1$ & $2.6\pm0.2$ & $7\pm3$ & $15\pm15$ & $6.0\pm0.5$ & 46 & 636/606 \\
               & \suz & 402050020 & $6.75\pm0.07$ & $4.6\pm0.7$ & $8.0\pm0.4$ & $27\pm2$ & $14\pm1$ &  81 & 628/606 \\
GX~349+2       & \suz & 400003010 & $6.97_{-0.01}$ & $3.9\pm0.2$ & $7.5\pm0.4$ & $26\pm1$ & $16\pm1$ & 105  & 1268/1106 \\
               & \suz & 400003020 & $6.64\pm0.03$ & $2.3\pm0.1$ & $10\pm2$ & $42\pm2$ & $9.2\pm0.6$ & 79  & 1262/1106\\
               & \xmm & 0506110101 & $6.76\pm0.01$ & $1.9\pm0.1$ & $6.0^{+0.5}$ & $55\pm2$ & $12.1\pm0.3$ & 79 & 2531/1831\\
Cyg~X-2      & \suz & 403063010 &  $6.97_{-0.02}$ & $6.5\pm1.3$ & $8.1\pm0.9$ & $25\pm3$ & $7.0\pm0.6$ & 65 & 1305/1129 \\
SAX~J1808.4$-$3658   & \suz & 903003010 &  $6.40^{+0.04}$ & $2.9\pm0.2$ & $12^{+7}_{-1}$ & $50^{+7}_{-2}$ & $0.70\pm0.06$ & 134 & 2987/2550\\
                     & \xmm & 0560180601 & $6.4^{+0.06}$ & $3.0\pm0.2$ & $13\pm4$ & $59\pm4$ & $0.68\pm0.04$ & 118  &  2124/1950 \\ 
HETE~J1900.1$-$2455  & \suz & 402016010 &  $6.97_{-0.2}$ & $5\pm1$ & $6^{+1}$   &  $20\pm4$ &  $0.16\pm0.02$ & 122 &   1146/1226\\  
\enddata
\tablecomments{Fe K line parameters from phenomenological modeling. All uncertainties are 1$\sigma$. Where no uncertainty is given, that parameter was fixed.  In the diskline model, the emissivity of the disk follows a power-law radial dependence, $R^{-q}$, where $q$ is the emissivity index we report here.  The normalization of the diskline is the line flux in photons cm$^{-2}$ s$^{-1}$. }
\label{tab:phenom_line}
\end{deluxetable*}

\begin{deluxetable*}{lclcccc}
\tablecolumns{7}
\tablewidth{0pc}
\tablecaption{Source fluxes and luminosities in the 0.5 -- 25 keV range}
\tablehead{Source & Mission & Obs. ID & \multicolumn{2}{c}{Phenomenological} & \multicolumn{2}{c}{Reflection} \\
 & & & Unabs. Flux  & Luminosity  &  Unabs. Flux  & Luminosity  \\
 & & & ($10^{-8}$ erg s$^{-1}$ cm$^{-2}$) & ($10^{37}$ erg s$^{-1}$) & ($10^{-8}$ erg s$^{-1}$ cm$^{-2}$) & ($10^{37}$ erg s$^{-1}$) }
\startdata
Serpens~X-1     & \suz & 401048010 & $1.19\pm0.01$ & $10.0\pm5.0$   & $1.32\pm0.08$ & $11.1\pm5.6$ \\
               & \xmm & 0084020401 & $0.70\pm0.01$ & $5.9\pm3.0$    & $0.71\pm0.01$ & $6.0\pm3.0$ \\ 
               & \xmm & 0084020501 & $0.59\pm0.01$ & $5.0\pm2.5$    & $0.60\pm0.01$ & $5.1\pm2.5$ \\
               & \xmm & 0084020601 & $0.65\pm0.01$ & $5.5\pm2.7$    & $0.66\pm0.01$ & $5.6\pm2.8$ \\ 
4U~1636$-$53   & \xmm & 0303250201 & $0.17\pm0.01$ & $0.73\pm0.05$  & $0.18\pm0.02$ & $0.78\pm0.09$ \\
               & \xmm & 0500350301 & $0.39\pm0.01$ & $1.7\pm0.1$    & $0.37\pm0.01$ & $1.6\pm0.1$ \\
               & \xmm & 0500350401 & $0.48\pm0.01$ & $2.1\pm0.1$    & $0.46\pm0.03$ & $2.0\pm0.1$ \\
4U~1705$-$44    & \suz & 401046010 & $0.17\pm0.01$ & $0.68\pm0.06$  & $0.57\pm0.05$ & $2.3\pm0.3$ \\
                & \suz & 401046020 & $0.61\pm0.01$ & $2.5\pm0.2$    & $0.49\pm0.01$ & $2.0\pm0.1$ \\
                & \suz & 401046030 & $0.30\pm0.01$ & $1.2\pm0.1$    & $0.31\pm0.01$ & $1.2\pm0.1$ \\
               & \xmm & 0402300201 & $0.043\pm0.001$& $0.17\pm0.01$  & $0.042\pm0.001$ &$0.17\pm0.01$ \\
4U~1820$-$30    & \suz & 401047010 & $1.21\pm0.01$ & $8.4\pm0.9$   & $1.21\pm0.13$ & $8.4\pm1.3$ \\
GX~17+2         & \suz & 402050010 & $2.37\pm0.01$ & $27.2\pm2.2$   & $2.27\pm0.01$ & $26.1\pm2.1$ \\
                & \suz & 402050020 & $2.60\pm0.01$ & $29.9\pm2.4$   & $2.62\pm0.01$ & $30.1\pm2.5$ \\
GX~349+2        & \suz & 400003010 & $2.35\pm0.03$ & $7.0\pm3.5$    & $2.36\pm0.13$ & $7.1\pm3.6$ \\
                & \suz & 400003020 & $1.97\pm0.01$ & $5.9\pm2.9$    & $1.92\pm0.01$ & $5.7\pm2.9$ \\
               & \xmm & 0506110101 & $2.38\pm0.01$ & $7.1\pm3.6$    & $2.40\pm0.09$ & $7.2\pm3.6$ \\
Cyg~X-2         & \suz & 403063010 & $2.22\pm0.01$ & $32.1\pm11.7$  & $2.30\pm0.01$ & $33.3\pm12.1$\\
SAX~J1808.4$-$3658 & \suz & 903003010 & $0.20\pm0.01$ & $0.29\pm0.02$ & $0.27\pm0.02$ & $0.40\pm0.04$\\ 
                  & \xmm & 0560180601 & $0.25\pm0.01$ & $0.37\pm0.03$ & $0.26\pm0.01$ & $0.38\pm0.03$\\
HETE~J1900.1$-$2455& \suz & 402016010 & $0.034\pm0.001$& $0.053\pm0.015$ & $0.032\pm0.005$ & $0.050\pm0.016$
\enddata
\tablecomments{Distances used are given in Table \ref{tab:obs}.}
\label{tab:fL}
\end{deluxetable*}

\begin{figure}
\centering
\includegraphics[width=7.5cm]{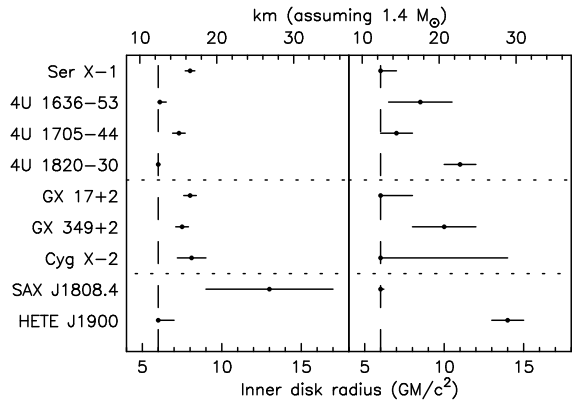}
\caption{Measured inner disk radii ($GM/c^2$) from phenomenological fits (left) and reflection fits (right). Where there are multiple observations, the one with the smallest fractional uncertainty was chosen.  The upper x-axis shows the inner disk radius in km assuming a 1.4 M$_\odot$ neutron star.  The horizontal dotted lines separate the Z sources (top), atolls (middle) and accreting millisecond X-ray pulsars (bottom). The vertical dashed line marks 6 $GM/c^2$, the lower limit on the inner disk radius allowed by the model.}
\label{fig:rin_range}
\end{figure}

The distribution of measured inner disk radii is shown in Fig.~\ref{fig:rin_range}.  We only use one measurement per source, choosing the one with the smallest fractional uncertainty.  The vast majority of the measured inner disk radii fall into a relatively narrow range of 6 -- 15 $GM/c^2$.  Only two observations (of Ser~X-1) fall outside this range, and even then the inner disk radii from those observations is only a little more than 1$\sigma$ away.   Note that there are a number of fits in which $R_{\rm in}$ pegs at the lower limit of the model (6 $GM/c^2$, the innermost stable circular orbit in the Schwarzschild metric).  This corresponds to about 12 km, assuming a 1.4~M$_\odot$ neutron star.  Given that this lower limit is very similar to predicted neutron star radii, smaller inner disk radii are not expected.    The lines in the observations with $R_{\rm in} = 6$ $GM/c^2$ are well fit by the model, and do not show large residuals in the red wing of the line (which would be expected if $R_{\rm in}$ were smaller than measured).

One possibility for the large number of sources that reach the lower limit of the model is that there is some contribution to the broadening of the line from Compton scattering.    \citet{reis09_1705} found that for 4U~1705$-$44 the measured inner disk radius increased when such effects were taken into account using reflection modeling.

We note that the differences in equivalent width (EW) between our original \suz{} analysis of Ser~X-1, GX~349+2 and 4U~1820$-$30 and  this analysis can be attributed to updated responses.  During the original analysis the standard response matrices had to be used, which did not fully account for the window mode, and therefore gave incorrect fluxes.

\subsection{Reflection models}
\label{sec:reflection}

As noted earlier, the Fe K$\alpha$ emission line is only the most prominent feature expected from disk reflection where hard ionizing flux irradiates the accretion disk leading to line emission.  In order to more self-consistently model the spectra, we fit reflection models which include line emission from important astrophysical metals (the strongest line being Fe K$\alpha$), as well as effects such as Compton broadening.

\begin{figure}
\centering
\includegraphics[width=7.5cm]{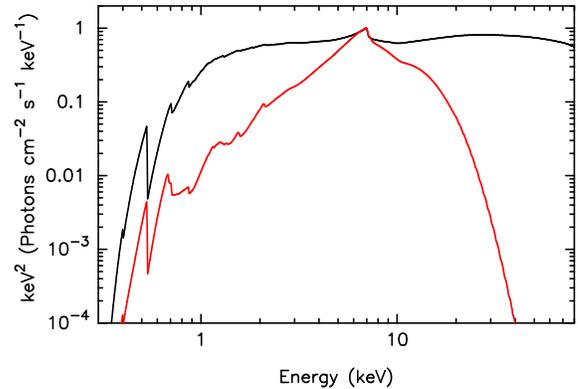}
\caption{Blurred reflection models for illumination by a power-law, spectral index = 2 (black) and for illumination by a blackbody, $kT = 2$ keV (red) both for an ionization parameter of 1000 \citep[models used are from][]{ballantyne01,ballantyne04}.  The models are normalized so that they both peak at 1.  The spectra are blurred using the rdblur model (a convolution with the diskline model), both assuming an  emissivity index $q = 3$, inner disk radius = 10 $GM/c^2$, outer disk radius = 1000  $GM/c^2$, and inclination = 30$^{\circ}$.  Both spectra are also absorbed assuming a quite typical value of $N_{\rm H} = 5 \times 10^{21}$ cm$^{-2}$.}
\label{fig:refl_mo}
\end{figure}

As is clear from the phenomenological modeling of the spectra \citep[and as has been noted before, e.g.,][]{cackett08,iaria09}, in the atoll and Z sources, the blackbody component (potentially associated with the boundary layer) is often dominating between 8 -- 20 keV. Thus, it is this emission that is providing the vast majority of ionizing flux that illuminates the accretion disk forming the Fe K line (and reflection spectrum).  In order to model the reflection component, one therefore needs to employ a reflection model where a blackbody component provides the illuminating flux \citep{ball_stroh04,ballantyne04}.   This differs from the standard reflection models used to model black hole sources \citep[e.g.][]{ballantyne01,ross07} which assume illumination by a power-law spectrum.  The resulting reflection spectra are quite different \citep[see Fig.\ref{fig:refl_mo}, and][for details]{ballantyne04}, with the blackbody reflection model dropping off quickly above 10 keV compared to the power-law reflection.

Here, we use models created to study the Fe K emission line observed in 4U~1820$-$30 during a superburst, where a constant-density slab is illuminated by a blackbody\footnote{The reflection models used here are to be made publically available on the XSPEC website} \citep{ball_stroh04,ballantyne04}.  We use the model assuming solar abundances for all sources except  4U~1820$-$30, where we use the model for a He system \citep{ball_stroh04}.  We note here that the density used in these models is $n_{\rm H} = 10^{15}$ cm$^{-3}$.  This is significantly lower than the density expected at the inner disk in neutron star systems.  \citet{ballantyne04} tested increasing the density and found that it did not significantly change the Fe K emission, though changes at lower energies ($<1$ keV) were more important.  While a higher density model would be more appropriate here, there is not yet one available for use.

We use a model where the irradiating blackbody component is included in the reflection model, in order to calculate the reflection fraction, $R$, which is given in Table~\ref{tab:reflect_line} with respect to that for a point source above a slab.  The overall model is given by a combination of the reflected spectrum and this blackbody.  In the model, the ionization parameter is defined as

\begin{equation}
\xi = \frac{4\pi F_{\rm x}}{n_{\rm H}}
\end{equation}

where, $F_{\rm x}$ is the irradiating flux (erg cm$^{-2}$ s$^{-1}$) in the 0.001 -- 100 keV range  and $n_{\rm H}$ is the hydrogen number density.  The ionization parameter, $\xi$, therefore has units erg cm s$^{-1}$. The normalization of the reflection model is simply a scale factor that sets the overall strength of the combined blackbody and reflection spectrum, whereas the reflection fraction sets the relative strength of the two.  The observed, unabsorbed flux (erg cm$^{-2}$ s$^{-1}$) of the blackbody plus reflection spectrum from the source in the 0.001 -- 100 keV range is therefore given by
\begin{equation}
F_{\rm obs, unabs} = {\rm Norm}\times F_{\rm x}(1+R)
\end{equation} 
where, Norm is the normalization of the model.  Rearranging gives
\begin{equation}
{\rm Norm} = \frac{4\pi F_{\rm obs, unabs}}{\xi n_{\rm H}(1+R)}
\end{equation}
and the density used in the models, $n_{\rm H} = 10^{15}$ cm$^{-3}$, is substituted.

To make these spectral fits more easily reproducible by other models we have calculated the equivalent blackbody normalization (using the definition of the XSPEC model bbody), from the best fitting value of the reflection model normalization and $\xi$. The strength of the reflection component relative to this is given by the reflection fraction.  So that the results are reproducible using the same model as used here, we also quote the normalization value directly from the model (as defined above).  

The strength of the Fe K line is dependent on both the ionization parameter and the temperature of the irradiating blackbody.  As the blackbody temperature increases, the number of photons above the Fe K edge increases, leading to a greater EW.  The EW peaks both at low ionization parameters, $\log \xi = 1.0$, where Fe K$\alpha$ is at 6.4 keV, and then at $\log \xi \sim 2.7$ where the line is at 6.7 keV.  At the peak for the higher ionization parameter, the EW is between $\sim 100$ -- 300 eV for a blackbody temperature in the range 1.5 keV to 3.0 keV.  This is shown clearly in figure 3 of Ballantyne (2004).  These equivalent widths are for a reflection fraction of 1.  Lower reflection fractions lead to smaller equivalent widths.

We relativistically blur this reflection model by convolving with the \verb!rdblur! model in XSPEC.  \verb!rdblur! is simply a convolution with the \verb!diskline! model, allowing the entire reflection component to be blurred by the relativistic effects present at the inner accretion disk. This blurred reflection component is then included in the model instead of the \verb!diskline! model used in the phenomenological fits.  This modeling acts to self-consistently model the entire reflection spectrum (multiple emission lines and continuum) rather than just treating the Fe K line alone.

\begin{figure}
\centering
\includegraphics[angle=-90, width=4cm]{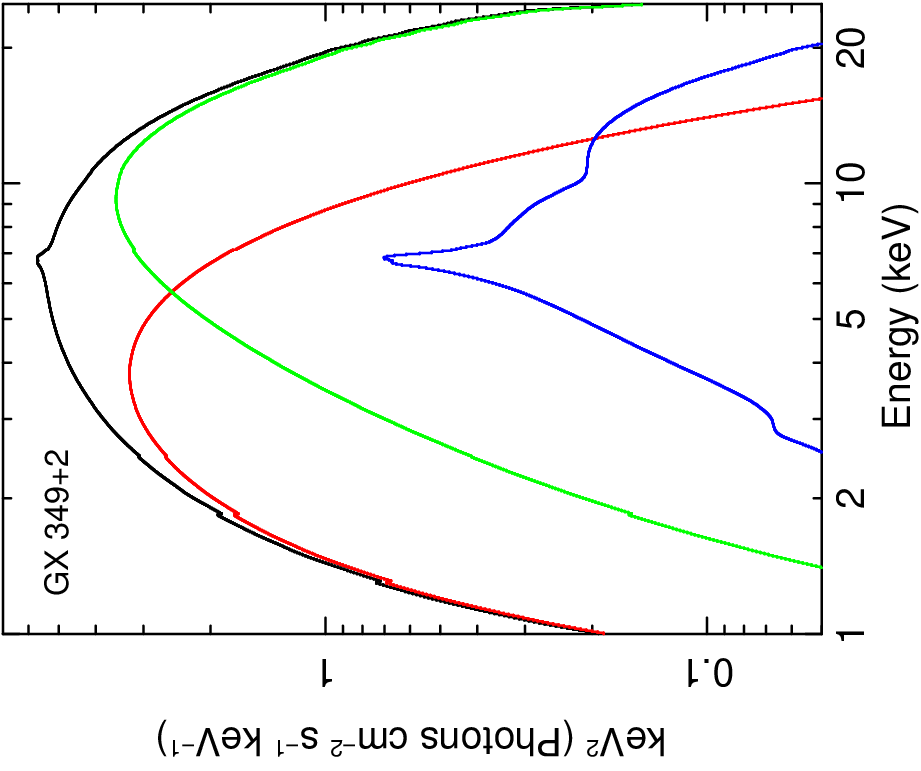}
\includegraphics[angle=-90, width=4cm]{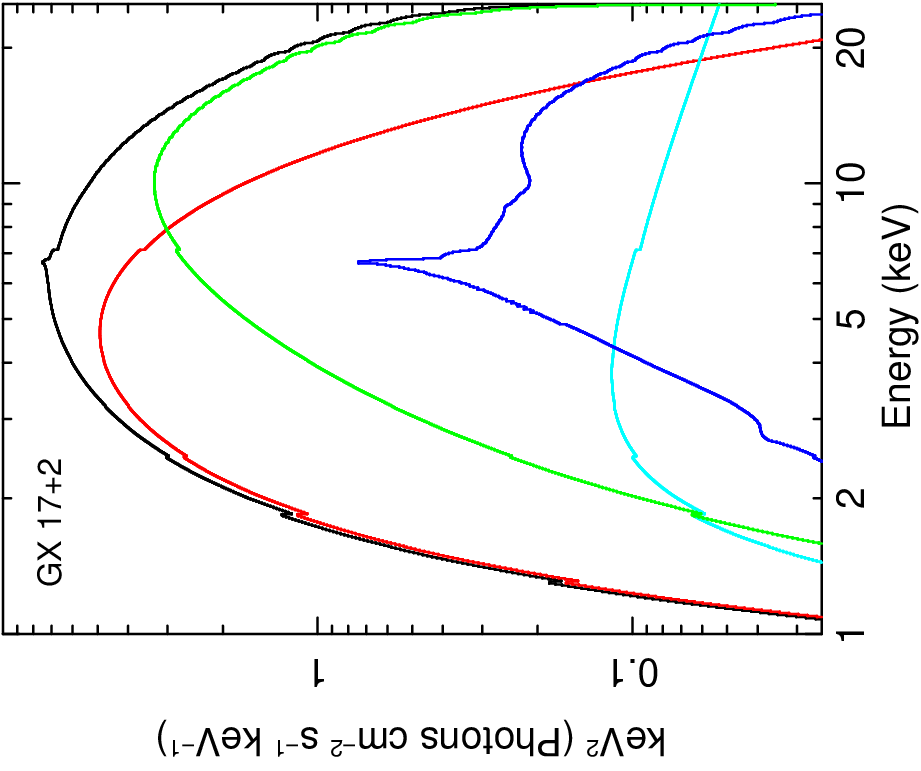}
\caption{Reflection model fit to the second \suz{} observation of GX~349+2 (left) and the second \suz{} observation of GX~17+2 (right).  The individual model components are shown - red: disk blackbody, green: blackbody, cyan: power-law, blue: reflection component.}
\label{fig:reflect}
\end{figure}

We are able to achieve good fits with this reflection modeling, generally finding similar inner disk radii as with phenomenological modeling (see spectral fit parameters in Tables \ref{tab:reflect_cont} and \ref{tab:reflect_line}).  An example of our reflection fits is shown in Fig. \ref{fig:reflect}.  It is important to note that because the blackbody component drops off over 8 -- 20 keV, so too does the reflection component.  The typical strong Compton hump seen in the black hole sources (both in AGN and X-ray binaries) is therefore not apparent because of this.

\tabletypesize{\tiny}
\begin{deluxetable*}{lclcccccccccc}
\tablecolumns{13}
\tablewidth{0pc}
\tablecaption{Reflection model parameters: continuum}
\setlength{\tabcolsep}{0.04in}
\tablehead{Source & Mission & Obs. ID & $N_H$ & \multicolumn{2}{c}{Disk Blackbody} & \multicolumn{2}{c}{Blackbody} & \multicolumn{2}{c}{Power-law}  & \multicolumn{3}{c}{Gaussian} \\
 & & & $10^{22}$ cm$^{-2}$ & $kT_{\rm in}$ (keV) & Norm. & $kT$ (keV) & Norm. ($10^{-2}$) & $\Gamma$ & Norm. & Line energy (keV) & $\sigma$ (keV) & Norm. $(10^{-2})$}
\startdata
Serpens~X-1  & \suz & 401048010 &  $0.75\pm0.02$ & $1.22\pm0.01$ & $110\pm5$ & $2.29\pm0.01$ &  $5.3\pm0.5$  & $3.80\pm0.11$ & $1.19\pm0.01$ & -- & -- & --  \\
               & \xmm & 0084020401 & $0.35\pm0.01$ & $0.97\pm0.01$ & $192\pm4$ & $1.79\pm0.01$ & $4.2\pm0.2$ & -- & -- & $1.09\pm0.01$ & $0.13\pm0.01$ & $1.4\pm0.1$  \\
               & \xmm & 0084020501 & $0.35\pm0.01$ & $1.01\pm0.01$ & $166\pm6$ & $1.88\pm0.01$ & $2.95\pm0.07$ & -- & -- & $1.10\pm0.01$ & $0.17\pm0.01$ & $1.7\pm0.1$   \\
               & \xmm & 0084020601 & $0.37\pm0.01$ & $0.99\pm0.01$ & $180\pm1$ & $1.82\pm0.01$ & $3.53\pm0.06$ & -- & -- & $0.95\pm0.01$ & $0.29\pm0.01$ & $4.6\pm0.1$  \\
4U~1636$-$53     & \xmm & 0303250201 & $0.37\pm0.01$ & $0.21\pm0.01$ & $10500\pm2000$ & $1.92\pm0.02$ & $0.06\pm0.02$  &  $1.94\pm0.06$ & $0.215\pm0.004$ & -- & -- & -- \\
               & \xmm & 0500350301 & $0.48\pm0.01$ & $0.93\pm0.01$ & $78\pm6$ & $2.13\pm0.04$  & $0.9\pm0.3$  & $3.5\pm0.1$ & $0.51\pm0.01$ & -- & -- & -- \\
               & \xmm & 0500350401 & $0.47\pm0.01$ & $1.00\pm0.01$ & $84\pm2$ & $2.10\pm0.01$  & $1.3\pm0.2$  & $3.5\pm0.1$ & $0.52\pm0.01$ & -- & -- & -- \\
4U~1705$-$44   & \suz & 401046010 &  $2.0$ & $0.08\pm0.01$ & $(2.6^{+3.4}_{-1.9})\times10^8$   &  $1.34\pm0.15$ &  $0.016^{+0.026}_{-0.006}$ & $1.68\pm0.01$ & $0.13\pm0.03$ & -- & -- & -- \\
               & \suz & 401046020 & $1.67\pm0.01$ & $1.04\pm0.01$ & $107\pm4$ & $2.42\pm0.02$  & $2.3\pm0.4$  & -- & -- & -- & -- & -- \\
               & \suz & 401046030 & $2.02\pm0.02$ & $0.81\pm0.01$ & $89\pm12$ & $2.12\pm0.01$  & $0.90\pm0.07$  & $3.02\pm0.04$ & $0.47\pm0.05$ & -- & -- & --  \\
               & \xmm & 0402300201 & $1.55\pm0.01$ & $1.42\pm0.01$ & $2.63\pm0.11$ & $3.15_{-0.05}$ & $0.22\pm0.06$ & -- & -- & -- & -- & --  \\
4U~1820$-$30   & \suz & 401047010  & $0.21\pm0.01$ & $1.14\pm0.01$ & $119\pm4$ & $2.50\pm0.04$ & $8.6\pm0.2$  & $2.38\pm0.05$ & $0.32\pm0.30$ & -- & -- & -- \\
GX~17+2        & \suz & 402050010 & $2.10\pm0.06$ & $1.69\pm0.02$ & $99\pm5$ & $2.65\pm0.02$ & $6.7\pm1.3$  &  $1.8\pm0.5$ & $0.017^{+0.211}_{-0.009}$ & -- & -- & -- \\
               & \suz & 402050020 & $2.15\pm0.05$ & $1.72\pm0.03$ & $95\pm4$ & $2.59\pm0.01$ & $9.4\pm1.3$  & $2.5\pm0.2$ & $0.27^{+0.29}_{-0.15}$ & -- & -- & -- \\
GX~349+2       & \suz & 400003010 & $0.77\pm0.01$ & $1.55\pm0.01$ & $109\pm4$ &	$2.33\pm0.01$ & $11.5\pm2.1$ &  -- & --  & -- & -- & -- \\
               & \suz & 400003020 & $0.80\pm0.01$ & $1.47\pm0.01$ & $112\pm4$ & $2.35\pm0.01$ & $8.9\pm0.8$  & -- & -- & -- & -- & -- \\
               & \xmm & 0506110101& $0.67\pm0.01$ & $1.24\pm0.01$ & $221\pm6$ & $1.89\pm0.01$ & $15.1\pm0.1$  & -- & -- & $1.07\pm0.01$ & $0.09\pm0.01$ & $3.4\pm0.1$  \\
Cyg~X-2      & \suz & 403063010 & $0.24\pm0.01$ & $1.50\pm0.01$ & $163\pm1$ & $2.23\pm0.01$ & $3.5\pm0.2$  & $2.93\pm0.04$ & $0.87\pm0.09$ & $1.00\pm0.01$ & $0.11\pm0.01$ & $10\pm1$  \\
SAX~J1808.4$-$3658   & \suz & 903003010 & $0.30\pm0.01$ & $0.41\pm0.01$ & $1370\pm120$ & $0.91\pm0.01$ & $0.05\pm0.01$ & $2.14\pm0.02$ & $0.27\pm0.01$ & -- & -- & --   \\
                     & \xmm & 0560180601 & $0.27\pm0.02$ & $0.48\pm0.01$ & $232\pm21$ &	$0.19\pm0.01$  & $0.6\pm0.1$ & $2.09\pm0.05$ & $0.31\pm0.01$ & -- & -- & -- \\ 
HETE~J1900.1$-$2455  & \suz & 402016010 & $0.09\pm0.01$ & -- &	-- &  -- & -- & $2.18\pm0.01$  & $0.07\pm0.01$  &  -- & -- & --  \\
\enddata
\tablecomments{Continuum spectral fitting parameters from reflection modeling. All uncertainties are 1$\sigma$. The $\chi^2$ values for the fits are those in Tab.~\ref{tab:reflect_line}. Where no uncertainty is given, that parameter was fixed. Normalization of the disk blackbody is in $R_{\rm inner}^2/(D10)^2\cos{i}$ where $R_{in}$ is the inner disk radius in km, $D10$ is the distance to the source in units of 10 kpc, and $i$ is the disk inclination.  The normalization of the blackbody component is defined as $L39/(D10)^2$, where $L39$ is the luminosity in units of $10^{39}$ erg s$^{-1}$.  Normalization of the power-law component gives the photons keV$^{-1}$ cm$^{-2}$ s$^{-1}$ at 1 keV.  Finally, the normalization of the Gaussian is the total number of photons cm$^{-2}$ s$^{-1}$ in the line. Note that the lower limit to the blackbody temperature in the reflection fits using the model of \citet{ballantyne04} (i.e. all except SAX~J1808.4$-$3658 and HETE~J1900.1$-$2455) is 1.0 keV, and the upper limit is 3.15 keV.}
\label{tab:reflect_cont}
\end{deluxetable*}

\tabletypesize{\tiny}
\begin{deluxetable*}{lclccccccc}
\tablecolumns{10}
\tablewidth{0pc}
\tablecaption{Reflection model parameters}
\tablehead{Source & Mission & Obs. ID & \multicolumn{6}{c}{Blurred reflection parameters} & $\chi^2$/dof \\
 & &  & Emissivity Index & $R_{\rm in}$ ($GM/c^2$) & Inclination $(^\circ)$ & $\log{\xi}$ & Norm. ($10^{-26}$) & Refl. fraction &}
\startdata
Serpens~X-1  & \suz & 401048010      & $2.2\pm0.1$ & $6^{+1}$ & $16\pm1$ & $2.6\pm0.1$ & $13.6\pm1.3$ & $0.18\pm0.01$ & 1526/1105 \\
               & \xmm & 0084020401   & $>6$ & $107\pm11$ & $5\pm2$ & $2.8\pm0.1$ & $7.5\pm0.3$ & $0.10\pm0.01$ & 2213/1831\\
               & \xmm & 0084020501   & $2.2\pm0.1$ & $15\pm2$ & $9\pm2$ & $2.6\pm0.1$ & $8.2\pm0.2$ & $0.16\pm0.01$ & 1982/1831 \\
               & \xmm & 0084020601   & $2.9\pm0.1$ & $44\pm2$ & $<3$ & $2.8\pm0.1$ & $6.4\pm0.1$ & $0.11\pm0.01$ & 2199/1831 \\
4U~1636$-$53     & \xmm & 0303250201 & $2.6\pm0.1$ & $8.5\pm2.0$ & $>68$ & $2.5\pm0.1$ & $0.20\pm0.05$ & $5.0_{-0.5}$ & 1881/1832 \\
               & \xmm & 0500350301   & $3.2\pm0.4$ & $13\pm4$ & $63_{-7}^{+16}$ & $3.2\pm0.1$ &	$0.54\pm0.20$  & $0.52\pm0.08$ & 2080/1832\\
               & \xmm & 0500350401   & $2.9\pm0.1$ & $13\pm2$ & $>75$ & $3.0\pm0.1$ & $1.3\pm0.2$  & $0.42\pm0.02$ & 2339/1832\\
4U~1705$-$44   & \suz & 401046010    & 3.0  & 10.0  &  24  & $3.0\pm0.2$ & $0.017^{+0.028}_{-0.007}$  & $5.0_{-2.5}$ & 1042/1127\\
               & \suz & 401046020    & $6.0\pm0.8$ & $7\pm1$ & $32\pm1$ & $3.0\pm0.1$ & $3.0\pm0.5$  &  $0.43\pm0.02$ & 1387/1124 \\
               & \suz & 401046030    & $5.5\pm2.2$ & $11\pm6$  & $<4$ & $2.0\pm0.1$ & $8.7\pm0.7$  & $1.1\pm0.1$ & 1135/1122\\
               & \xmm & 0402300201   & $>3.4$ & $34_{-12}^{+5}$ & $16\pm3$ & $3.0\pm0.1$ & $0.26\pm0.07$  & $0.21\pm0.02$ &  1766/1752\\
4U~1820$-$30   & \suz & 401047010    & $>6$ & $11\pm1$ & $27\pm2$ & $2.6\pm0.1$ & $28.1\pm0.5$  & $0.11\pm0.02$ & 1336/1110\\
GX~17+2        & \suz & 402050010    & $2.2\pm0.1$ & $6^{+2}$ & $<11$ & $2.7\pm0.1$ & $15.5\pm2.9$   &  $0.20\pm0.03$ & 647/606\\
               & \suz & 402050020    & $2.3\pm0.3$ & $6^{+2}$ & $<6$ & $2.7\pm0.1$ & $20.0\pm2.7$  & $0.21\pm0.02$ & 684/606\\
GX~349+2       & \suz & 400003010    & $2.4\pm0.1$ & $10\pm2$ & $28\pm1$ & $2.9\pm0.1$ & $15.0\pm2.7$  & $0.24\pm0.01$ & 1341/1109  \\
               & \suz & 400003020    & $>1.8$ & $386\pm97$ & $18\pm2$ & $3.1\pm0.1$ & $7.7\pm0.7$  & $0.22\pm0.01$ & 1276/1109\\
               & \xmm & 0506110101   & $0.3\pm0.4$ & $6^{+54}$ & $48\pm2$ & $2.9\pm0.1$ & $22.3\pm0.1$  & $0.14\pm0.01$ & 2500/1831 \\
Cyg~X-2      & \suz & 403063010    & $2.0\pm0.1$ & $6^{+8}$ & $<9$ & $2.6\pm0.1$ & $9.5\pm0.4$  & $0.31\pm0.03$ & 1379/1135\\
SAX~J1808.4$-$3658   & \suz & 903003010   & $>5.2$ & $14\pm2$ & $67_{-4}$ & $2.7\pm0.1$ & $10.2\pm0.5$  & $0.86\pm0.08$ & 581/620 \\
                     & \xmm & 0560180601  & $2.2\pm0.1$ & $6.0^{+0.2}$ & $67_{-1}$ & $2.8\pm0.1$  & $8.6\pm0.2$  & $0.34\pm0.03$ & 2120/1952\\ 
HETE~J1900.1$-$2455  & \suz & 402016010   & $5.2\pm0.8$ & $14\pm1$   &  $<4$ &  $2.8\pm0.1$ & $2.1\pm0.3$ & $0.31\pm0.03$ & 1101/1227
\enddata
\tablecomments{All uncertainties are 1$\sigma$. Where no uncertainty is given, that parameter was fixed. The normalization of the reflection model (Norm.) scales the overall flux of this component.  Please see section~\ref{sec:reflection} for further discussion.}
\label{tab:reflect_line}
\end{deluxetable*}

The two sources where we do not fit this blackbody reflection model are the accreting millisecond pulsars (SAX~J1808.4$-$3658 and HETE~J1900.1$-$2455).  Both sources show particularly hard spectra, where the high energy continuum is dominated by a power-law.  We therefore fit these spectra with the constant density ionized disk reflection model \citep{ballantyne01}, where a power-law irradiates a slab. As with the blackbody reflection model, the model normalization is dependent on $\xi$, and is given by the same definition as above.  We therefore have calculated the equivalent power-law normalization at 1 keV so that these results are reproducible using other reflection models.  Again, we also report the best-fitting normalization directly from the model so that these results can be easily reproduced using the same model.  This model fits the data well in both cases.  HETE~J1900.1$-$2455 requires only the power-law reflection component, and no other continuum components are needed.  SAX~J1808.4$-$3658, on the other hand, requires both a disk blackbody and blackbody in addition to the power-law reflection component.

All spectral parameters from this reflection fitting are given in Tables ~\ref{tab:reflect_cont} and \ref{tab:reflect_line}, and source fluxes and luminosities are given in Table~\ref{tab:fL}.  We also show the range of measured inner disk radius from the reflection component in Fig. \ref{fig:rin_range}.  We find a large number of source have their measured inner disk radii pegged at the model limit of 6 $GM/c^2$, we discuss this in more detail later. 

\section{Simultaneous Timing Observations}

Fe K emission lines have been detected simultaneously with kHz QPOs previously \citep[e.g. GX~17+2, ][]{cackett_chazss_09}, however, there has not yet been the sensitivity to test whether the line and kHz QPO originate from the same region.  While simultaneity is clearly important due to known changes in kHz QPO frequency with source state, comparisons of non-simultaneous measurements of $R_{\rm in}$ from Fe K lines and kHz QPOs are promising \citep{cackett08}.  Simultaneous observations were performed by the {\it RXTE} Proportional Counter Array for both the \suz{} pointings of GX~17+2 and the second Cyg~X-2 observations. Both sources are known kHz QPO sources.  Those QPOs were previously reported by  \cite{wijnands97} and \cite{homan02} for GX~17+2, and by \cite{wijnands98} for Cyg~X-2.

We have analyzed the PCA observations, considering segments of continuous observation (ObsIDs): 2 and 5 ObsIDs worth of data were recorded for  GX~17+2 and Cyg~X-2 respectively. For each ObsID, we have computed an average Power Density Spectrum (PDS) with a 1 Hz resolution, using events recorded between 2 and 40 keV. The PDS are normalized according to \citet{leahy83}, so that in the absence of significant deadtime (as is the case here for both sources), the Poisson noise level is flat and constant with a value close to 2. The PDS so computed was then blindly searched for excess power between 300 Hz and 1400 Hz using a scanning technique, as presented in \citet{boirin00}. We failed to detect any kHz QPOs in any of the observations, for both sources. The analysis was also performed considering photons of energy between 5 and 40 keV, but no kHz QPOs were detected. Finally, we have also produced an averaged PDS containing all observations. No QPOs were detected either.

We have derived upper limits for QPOs of various widths, as listed in Table \ref{tab:timing} for each pointed observation. Using data from combined observations, the upper limits are 15.5\% and 17.1\% for GX~17+2, and 4.6\% and 5.1\% for Cyg~X-2 for signal of 100 and 150 Hz width respectively. These upper limits are not constraining for GX~17+2, whose lower and upper kHz QPOs have maximum RMS amplitudes of $\sim 4$\% and $\sim 6$\% respectively \citep{wijnands97,homan02}. This is explained by the combination of a very low source count rate and a reduced number of operating PCA units (2 or 3). For Cyg~X-2, which displays larger source count rates, the RMS upper limits are significantly lower, but unfortunately still above the maximum RMS amplitude of its kHz QPO reported so far \citep{wijnands98}.

\tabletypesize{\scriptsize}
\begin{deluxetable*}{ccccccccccc}
\tablecolumns{11}
\tablewidth{0pc}
\tablecaption{{\it RXTE} observations of GX~17+2, Cyg~X-2 and HETE~J1900.1$-$2455}
\tablehead{Name & ObsID & Date & Time & T$_{\rm obs}$ & Rate & Bkg & N$_{\rm PCU}$ & RMS$_{50}$ (\%)& RMS$_{100}$ (\%) & RMS$_{150}$ (\%)}
\startdata
    GX~17+2 &    90022-07-06-01&19/09/2007&09:11&$2944.0$&$79.5$&$30.0$&$2$&$23.3$&$28.7$&$33.0$\\
    GX~17+2 &90022-07-07-00&27/09/2007&15:15&$1392.0$&$168.1$&$53.7$&$3$&$17.5$&$19.6$&$21.3$\\
    Cyg~X-2 & 93443-01-01-03&01/07/2008&01:34&$2016.0$&$644.3$&$40.8$&$2$&$6.8$&$8.3$&$9.4$\\
    Cyg~X-2 & 93443-01-01-04&01/07/2008&03:12&$1776.0$&$602.1$&$37.0$&$2$&$5.8$&$6.7$&$7.6$\\
    Cyg~X-2 & 93443-01-01-05&01/07/2008&04:50&$1536.0$&$464.2$&$39.3$&$2$&$8.1$&$9.5$&$10.5$\\
    Cyg~X-2 & 93443-01-01-06&01/07/2008&06:28&$1408.0$&$462.6$&$37.4$&$2$&$6.6$&$8.0$&$9.0$\\
    Cyg~X-2 & 93443-01-01-08&01/07/2008&11:09&$1344.0$&$772.0$&$16.5$&$2$&$6.0$&$7.0$&$7.9$ \\
    HETE~J1900.1$-$2455 &93030-01-16-00&13/10/2007&12:49&$3392.0$&$88.3$&$55.2$&$3.0$&$27.8$&$34.9$&$40.8$    \\
    HETE~J1900.1$-$2455 &93030-01-17-00&21/10/2007&10:52&$3392.0$&$120.3$&$54.0$&$3.0$&$19.6$&$24.1$&$27.4$    
\enddata
\tablecomments{T$_{\rm obs}$ is the observation duration (s) and Rate and Bkg give the source and background count rates (c/s) respectively. The number of active PCA units (N$_{\rm PCU}$) are listed for each ObsID. The upper limits on the QPO RMS amplitude (\%) have been computed for QPO width of 50, 100 and 150 Hz (RMS$_{50}$, RMS$_{100}$, RMS$_{150}$ respectively). They have been computed at the $3\sigma$ confidence level, meaning that the RMS listed would enable the detection of any QPOs with a $3\sigma$ accuracy on its amplitude (this corresponds to a $\sim 6 \sigma$ excess (single trial) in the power density spectrum, see \cite{boutelier09} for details).}
\label{tab:timing}
\end{deluxetable*}

Nevertheless for Cyg~X-2, inspecting lower frequencies, features around 5 and 50 Hz are visible in the ObsID averaged PDS. These are likely normal branch and horizontal branch QPOs, previously known from the source. It has been shown in sources like Cyg~X-2 \citep[the prototype being GX~17+2,][]{wijnands97,homan02} that the upper kHz QPO frequency correlates with the HBO frequency up to a certain frequency where the upper QPO frequency keeps increasing while the HBO one saturates \citep[see for instance Figure 5 in][]{psaltis99}. In Cyg~X-2, the HBO frequency has been shown to increase from about 35 Hz up to a saturating frequency of 55 Hz. At the same time, the upper QPO frequency varied from $\sim 730$ Hz to $\sim 1000$ Hz, while the corresponding HBO frequency ranged from 35 to 55 Hz \citep{wijnands98}. According to \citet{psaltis99} (their fig. 8), before the saturation, a power-law relationship exists between the HBO and upper QPO frequencies, in the form of $\rm \nu_{\rm HBO}=13.2 a_1 (\nu_{\rm upper}/1000)^{b_1}$ Hz, with $a_1\sim4.3$ and $b_1\sim1.4$ in the case of Cyg~X-2. The HBO QPO significance is $\sim 5 \sigma$ significance (single trial). Fitting the 1 -- 2048 Hz PDS with two Lorentzians and a constant to account for the Poisson noise level, we have estimated the HBO frequency to be $49.5^{+2.5}_{-2.4}$ Hz, the QPO width as  $7.0^{+4.3}_{-4.1}$ Hz for an RMS amplitude of about $3.5^{+0.7}_{-1.5}$\%. The latter two parameters are fully consistent with previous detections \citep{wijnands98}. Despite its low significance, it is therefore tempting to use the $\sim 50$ Hz HBO reported here (which is below the saturating frequency) to estimate with the above formula, the frequency of the upper QPO, that we failed to detect. This yields a frequency of about 910 Hz.

As we noted earlier, if both the upper kHz QPO and Fe K emission line arise from the same region in the disk, and the kHz QPO is associated with an orbital frequency, they can be combined to estimate the neutron star mass \citep{piraino00,cackett08}.  If we tentatively use the inferred 910 Hz from scaling of the HBO frequency, along with the inner disk radius from the iron line we get a mass estimate in the range 1.5 -- 2.4 M$_\odot$ (depending on whether we use the phenomenological or reflection modeling results), consistent with the optical mass measurement \citep{orosz99, elebert09}. This implies a consistent radius between the Fe K emission line and the kHz QPO.  But, again, we are extremely cautious about this given the large uncertainty in scaling the QPO frequency, and the possible range of inner disk radius.

Looking at the {\it RXTE} archive, two observations of HETE~J1900.1$-$2455 were performed contemporaneously with the \suz{} pointings of this source. Unfortunately, no kHz QPOs were detected. Upper limits are also listed in Table \ref{tab:timing}, but they are not very constraining on the presence of kHz QPOs due to the low source count rate and a comparatively large background.

Archival {\it RXTE} observations are also available for the three \xmm{} pointings of 4U~1636$-$53: the first observation was analyzed in \citet{barret05} and the latter two are also considered in \citet{boutelier10}.  We analyze the data using the same method as described above. As strong kHz QPOs were detected in many pointed observations, the QPO parameters (mainly the quality factor) were corrected for frequency drift using the method described in \citet{boutelier09}. A total of 10 observations were considered. In 4 of those, we detect twin kHz QPOs, in one we detect a single upper kHz QPO at $\sim 500$ Hz (i.e. the lower kHz QPO which is expected at much lower frequencies has merged with the noise and is no longer a QPO), and in two, a single lower kHz QPO \citep[for an identification of single kHz QPO, based on their quality factor and RMS amplitude see,][]{barret06}. The non detection of the upper kHz QPO in the latter two pointed observations is due to a lack of sensitivity, see \citet{boutelier10}, but the frequency of the upper kHz QPO can be approximated by the nearly linear function that links the two frequencies \citep[e.g.][]{belloni05}. The results of the fits are listed in Table~\ref{tab:1636}. 

\begin{deluxetable*}{cccccccccccc}
\tablecolumns{12}
\tablewidth{0pc}
\tablecaption{{\it RXTE} observations of 4U~1636$-$536 simultaneous with the three \xmm{} observations}
\tablehead{Name & ObsID & Date & Time & T$_{\rm obs}$ & Rate & Bkg & N$_{\rm PCU}$ &$\nu$ & FWHM & RMS (\%) & $R$}
\startdata
    4U~1636$-$53&70034-01-01-01&29/08/2005&03:24&$4338.0$&$406.9$&$53.3$&3&$509.6\pm35.6$&$191.4\pm85.8$&$12.3\pm2.0$&$3.0$
    \\
    4U~1636$-$53&91027-01-01-000&29/08/2005&16:35&$4434.0$&$420.6$&$78.8$&3&...&...&...&...
    \\
    4U~1636$-$53&91027-01-01-000&29/08/2005&18:06&$23362.0$&$347.4$&$78.9$&3&...&...&...&...
    \\
    4U~1636$-$53&93091-01-01-000&28/09/2007&14:47&$4464.0$&$482.7$&$40.6$&3&$643.5\pm2.5$&$17.8\pm8.3$&$6.5\pm1.0$&$3.4$\\
     & & & & & & & &$984.1\pm19.2$&$165.3\pm67.8$&$11.9\pm1.6$&$3.8$
    \\
    4U~1636$-$53&93091-01-01-000&28/09/2007&16:17&$23392.0$&$415.7$&$36.8$&2&$589.7\pm10.5$&$94.6\pm31.8$&$7.8\pm0.9$&$4.2$\\
     & & & & & & & &$906.3\pm5.9$&$88.9\pm15.8$&$10.0\pm0.6$&$7.7$
    \\
    4U~1636$-$53&93091-01-01-00&28/09/2007&22:47&$7104.0$&$179.5$&$17.1$&1&...&...&...&...
    \\
    4U~1636$-$53&93091-01-02-000&27/02/2008&03:46&$1280.0$&$690.2$&$58.3$&3&$645.9\pm5.7$&$36.5\pm15.8$&$8.4\pm1.2$&$3.4$
    \\
    4U~1636$-$53&93091-01-02-000&27/02/2008&04:35&$4352.0$&$688.6$&$34.9$&3&$646.0\pm2.6$&$19.3\pm7.4$&$5.5\pm0.7$&$3.8$\\
     & & & & & & & &$979.5\pm27.1$&$184.9\pm71.8$&$9.3\pm1.3$&$3.6$
    \\
    4U~1636$-$53&93091-01-02-000&27/02/2008&06:07&$20352.0$&$546.4$&$31.0$&2&$701.6\pm3.7$&$86.2\pm13.5$&$10.5\pm0.6$&$8.5$\\
     & & & & & & & &$1018.8\pm32.8$&$221.3\pm111.4$&$7.8\pm1.4$&$2.8$
    \\
    4U~1636$-$53&93091-01-02-00&27/02/2008&11:46&$10096.0$&$252.5$&$17.0$&1&$882.0\pm2.6$&$15.8\pm5.8$&$6.7\pm0.9$&$3.7$
\enddata  
\tablecomments{T$_{\rm obs}$ is the observation duration (s) and Rate and Bkg give the source and background count rates respectively. The number of active PCA units (N$_{\rm PCU}$) are listed for each ObsID. The frequency, Full Width Half Maximum, and RMS amplitude (\%) of each kHz QPO are reported with their $1\sigma$ error. $R$ is the ratio of the Lorentzian amplitude to its 1$\sigma$ error. An $R$ of 3 corresponds roughly to 6$\sigma$ excess power in the Fourier PDS, see \cite{boutelier09} for details. When two QPOs are detected in one ObsID, the first one refers to the lower kHz QPO, whereas the second one refers to the upper kHz QPO.}
\label{tab:1636}
\end{deluxetable*}

These simultaneous \xmm{} and {\it RXTE} observations are discussed extensively in \citet{altamirano09}; here we briefly discuss the consistency of our spectral fitting with the kHz QPO properties.  As \citet{altamirano09} find, when assuming a 1.4 M$_\odot$ neutron star, the inner disk radius (as measured by the Fe K line from phenomenological fits) disagrees with the implied inner disk radius assuming that the upper kHz QPO is a Keplerian frequency in the disk.  If instead one uses the phenomenological $R_{\rm in}$ measurements combined with the upper kHz QPO frequency to estimate a mass, the implied mass would be $4.2\pm0.5$ M$_\odot$, $1.7\pm0.2$ M$_\odot$ and $2.2\pm0.1$ M$_\odot$ from the three observations respectively (we use the kHz QPO frequencies from the {\it RXTE} observations that overlap with the \xmm{} observations for the longest period).  Clearly, the first observation estimates far too high a neutron star mass, while the other two observations give a more reasonable (though quite high) estimate. Note a high mass is predicted for 4U~1636$-$53 from the drop in kHz QPO coherence \citep{barret06}.  From reflection modeling, we find slightly different $R_{\rm in}$ measurements than for the phenomenological fits.  The corresponding mass estimates from reflection fits are: $2.6\pm1.0$ M$_\odot$, $0.8\pm0.4$ M$_\odot$and $0.7\pm0.2$ M$_\odot$.  Again, the separate measurements are not all consistent with each other, and this time two of the masses are clearly on the low side. 

Looking at the trends, if the two phenomena originate in the same part of the disk, a lower kHz QPO frequency should correspond to a larger inner disk radius (as measured by the Fe K line).  However, in 4U~1636$-$53 the opposite behavior is seen -- as the kHz QPO frequency increases there is not a corresponding decrease in $R_{\rm in}$, in fact, when using the $R_{\rm in}$ measurements from reflection modeling $R_{\rm in}$ is actually seen to increase.  This, however, is only based on three observations, and further, higher quality, observations are needed to make any strong conclusions.

\section{Discussion}
\label{discussion}

We have presented a comprehensive study of Fe K emission lines in 10 neutron star LMXBs.  In all cases, the emission line profiles are well modeled by a relativistic disk line assuming a Schwarzschild metric.  However, in one source, GX~340+0, the exact line profile is uncertain: a relativistic emission line fits the data well but only if either the abundance of Fe is non-standard, or alternatively there is an \ion{Fe}{26} absorption line superposed on the emission line.  Fitting of broadband spectra, makes it clear that the hot blackbody component dominates the spectrum from around 7 -- 20 keV (in the Z and atoll sources), and thus provides the vast majority of the flux irradiating the accretion disk and leading to the Fe K emission line.  We fit reflection models to the spectra, finding that in the Z and atoll sources a reflection model where the irradiating source is a blackbody does fit the spectra well.  In the accreting millisecond pulsars, however, a power-law dominates above 10 keV, and standard reflection models used to fit black hole spectra work well there.

\subsection{Inner disk radii}

The inner disk radii that we measure mostly fall into a small range: 6 -- 15 $GM/c^2$. Also consistent with this range is a recent independent measurement of the inner disk radius in 4U~1705$-$44 \citep[$R_{\rm in} = 14\pm2$;][]{disalvo09}. Where there are multiple observations of a source, in most cases there is not a large change in inner disk radius, which may suggest the inner disk radius does not vary with state.  While we note this from observations at different epochs, similar results have been found previously when looking at how the line profile changes with source state during one observation \citep{dai09,iaria09}.  

\begin{figure}
\centering
\includegraphics[width=7.5cm]{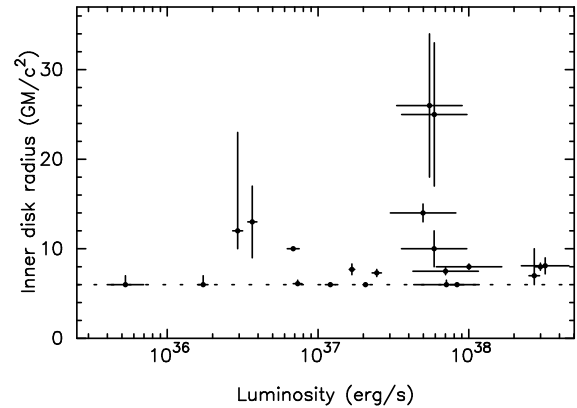}
\caption{Measured inner disk radii from phenomenological fits as a function of 0.5-25 keV source luminosity.  The dotted line indicates 6 $GM/c^2$, the smallest allowed radius in the model.}
\label{fig:lum_rin}
\end{figure}

The only source where we see variations in inner disk radii is Ser~X-1, and this is easily apparent when one compares the line profile observed by \suz{} with one of the \xmm{} observations (see Fig.~\ref{fig:diskline}).  Comparing these observations, the spectrum from the \suz{} observation of Ser~X-1 is harder than the \xmm{} observations -- although both the blackbody and disk blackbody components are hotter, the relative normalization of the blackbody component relative to the disk blackbody component is higher.  The ionizing flux during the \suz{} observation is therefore stronger, which would explain the higher line flux.   In all cases, though, the emitting radius inferred from the blackbody component is very similar.  The 0.5 -- 10 keV source flux is also highest during the \suz{} observation ($10.3\times10^{- 9}$ erg cm$^{-2}$ s$^{-1}$ compared to $6.3\times10^{- 9}, 4.4\times10^{- 9}$ and $4.9\times10^{- 9}$  erg cm$^{-2}$ s$^{-1}$ for the \xmm{} observations).  This may tentatively suggest that the disk extends closer to the stellar surface at the highest luminosities, though this is not confirmed when looking at all sources (Fig.~\ref{fig:lum_rin}).

We have also looked to see if there is any significant change in inner disk radius with 0.5 -- 25 keV luminosity for all sources (see Fig.~\ref{fig:lum_rin}).  There is no apparent systematic trend with luminosity, and the inner disk radius is found to be close to 6 $GM/c^2$ for the majority of sources/observations over almost 3 orders of magnitude in luminosity.  Similarly, we see no clear difference between atolls, Z sources and accreting millisecond pulsars, as can be seen in Fig.~\ref{fig:rin_range}. 

There is typically consistency between the inner disk radii measured from phenomenological and reflection spectral fits.  Nevertheless, in both cases many of the inner disk radii measurements are pegged at the lower limit of the model (the innermost stable circular orbit, ISCO).  In 4U~1705$-$44, \cite{reis09_1705} found that fitting with a model for a Kerr metric \citep[relevant for maximally spinning black holes][]{laor91} gave inner disk radii much smaller than expected neutron star radii. They suggested that as the disks in these systems are ionized one must also take broadening due to Comptonization into account by using relativistic ionized reflection models.  When they did this, the inner radius measured was significantly higher, and no longer smaller than expected stellar radii.  The reflection models employed here also take Compton broadening into account, yet, we still find that the inner disk radius pegs at the ISCO in a number of sources.

This may not necessarily be a problem. As neutron stars are spinning, their angular momentum will change the radius of the ISCO making it slightly smaller for a corotating disk \citep[see fig 1(a) of][to see how the ISCO changes with angular momentum]{miller07}.  Potentially, one could then measure the dimensionless angular momentum parameter, under the assumption that the inner disk is truncated at the ISCO and not at the boundary layer or stellar surface.  To test this, we fit a line profile with a variable angular momentum parameter \citep{brenneman06} to several of the best datasets in our sample.  However, in each case we found that this parameter was not well constrained, being consistent with zero.  Better, more sensitive data may be able to constrain this further.

We can also compare the inner disk radii from the Fe line fitting with that implied from the disk blackbody fits.  The normalization for the disk blackbody used here (\verb|diskbb|) is defined as $N = R_{\rm inner}^2 \cos{i}/(D10)^2$, where $R_{\rm inner}$ is the inner disk radius in km, $D10$ is the source distance in units of 10 kpc and $i$ the disk inclination.  Therefore, from the fitted normalization and the inclination determined from the Fe line fits, we can calculate the inner disk radius in km.  Assuming a reasonable range for the neutron star mass (here we use 1.4 -- 2.0 M$_\odot$) we can convert to units of $GM/c^2$ for direct comparison with the Fe line fits.  The disk radii determined this way are given in Table~\ref{tab:rin_diskbb}.   Where only an lower/upper limit for the inclination is found from the Fe line fits, we just use this limit for the inclination in the calculation.

\tabletypesize{\scriptsize}
\begin{deluxetable*}{lclcccccc}
\tablecolumns{9}
\tablewidth{0pc}
\tablecaption{Disk blackbody inner radius}
\tablehead{Source & Mission & Obs. ID & \multicolumn{3}{c}{Phenomenological} & \multicolumn{3}{c}{Reflection} \\
       &  &  & $R_{\rm inner}$ (km) & $R_{\rm in}$ (GM/c$^2$) & $R_{\rm in}$ (GM/c$^2$) & $R_{\rm inner}$ (km) & $R_{\rm in}$ (GM/c$^2$) & $R_{\rm in}$ (GM/c$^2$)  \\
      & & & & for 1.4 M$_\odot$ & for 2 M$_\odot$  & & for 1.4 M$_\odot$  & for 2 M$_\odot$ }
\startdata     
Serpens~X-1    & \suz & 401048010  &  $8.7\pm2.2$   & $4.2\pm1.0$      & $2.9\pm0.7$       &  $9.0\pm2.3$   & $4.4\pm1.1$      & $3.0\pm0.8$ \\
               & \xmm & 0084020401 &  $13.1\pm3.2$  & $6.3\pm1.6$      & $4.4\pm1.1$       &  $11.7\pm2.9$  & $5.6\pm1.4$      & $4.0\pm1.0$ \\
               & \xmm & 0084020501 &  $11.3\pm2.8$  & $5.5\pm1.4$      & $3.8\pm1.0$       &  $10.9\pm2.7$  & $5.3\pm1.3$      & $3.7\pm0.9$ \\
               & \xmm & 0084020601 &  $12.7\pm3.2$  & $6.1\pm1.5$      & $4.3\pm1.1$       &  $11.3\pm2.8$  & $5.5\pm1.4$      & $3.8\pm1.0$ \\
4U~1636$-$53   & \xmm & 0303250201 &  $143 \pm72$   & $70 \pm35$       & $49\pm25$         &  $100\pm49$     & $49\pm24$        & $34\pm17$   \\
               & \xmm & 0500350301 &  $12.7\pm6.3$  & $6.2\pm3.0$      & $4.3\pm2.1$       &  $7.9\pm2.2$   & $3.8\pm1.1$      & $2.7\pm0.7$ \\
               & \xmm & 0500350401 &  $18.8\pm9.4$  & $9.1\pm4.6$      & $6.4\pm3.2$       &  $10.8\pm5.3$   & $5.2\pm2.6$      & $3.7\pm1.8$ \\
4U~1705$-$44   & \suz & 401046010  &  $2910^{+4560}_{-1020}$ & $1400^{+2200}_{-490}$ & $990^{+1540}_{-340}$  &  $9780^{+6410}_{-3590}$  & $4740^{+3100}_{-1740}$ & $3320^{+2170}_{-1220}$\\
               & \suz & 401046020  &  $4.5\pm0.2$   & $2.2\pm0.1$      & $1.5\pm0.1$       &  $6.5\pm0.3$   & $3.2\pm0.1$      & $2.2\pm0.1$ \\
               & \suz & 401046030  &  $5.3\pm0.2$   & $2.6\pm0.1$      & $1.8\pm0.1$       &  $5.5\pm0.4$   & $2.7\pm0.2$      & $1.9\pm0.1$ \\
               & \xmm & 0402300201 &  $1.7\pm0.9$   & $0.8\pm0.4$      & $0.6\pm0.3$       &  $0.96\pm0.04$   & $0.47\pm0.02$  & $0.33\pm0.01$ \\
4U~1820$-$30   & \suz & 401047010  &  $7.8\pm0.5$   & $3.8\pm0.2$      & $2.6\pm0.2$       &  $8.8\pm0.5$   & $4.3\pm0.2$      & $3.0\pm0.2$ \\
GX~17+2        & \suz & 402050010  &  $9.2\pm0.6$   & $4.5\pm0.3$      & $3.1\pm0.2$       &  $9.8\pm0.5$   & $4.8\pm0.2$      & $3.3\pm0.2$ \\
               & \suz & 402050020  &  $10.0\pm0.5$   & $4.8\pm0.2$      & $3.4\pm0.2$       &  $9.6\pm0.4$   & $4.6\pm0.2$      & $3.2\pm0.2$ \\
GX~349+2       & \suz & 400003010  &  $5.6\pm1.4$   & $2.7\pm0.7$      & $1.9\pm0.5$       &  $5.6\pm1.4$   & $2.7\pm0.7$      & $1.9\pm0.5$ \\
               & \suz & 400003020  &  $5.9\pm1.5$   & $2.8\pm0.7$      & $2.0\pm0.5$       &  $5.4\pm1.4$   & $2.6\pm0.7$      & $1.8\pm0.5$ \\
               & \xmm & 0506110101 &  $10.6\pm2.7$  & $5.1\pm1.3$      & $3.6\pm0.9$       &  $9.1\pm2.3$   & $4.4\pm1.1$      & $3.1\pm0.8$ \\
Cyg~X-2        & \suz & 403063010  &  $14.8\pm2.7$  & $7.2\pm1.3$      & $5.0\pm0.9$       &  $14.1\pm2.6$  & $6.8\pm1.2$      & $4.8\pm0.9$ \\
SAX~J1808      & \suz & 903003010  &  $1.1\pm0.1$   & $0.51\pm0.04$    & $0.36\pm0.03$      &  $20.7\pm2.0$  & $10.0\pm1.0$      & $7.0\pm0.7$\\
               & \xmm & 0560180601 &  $70^{+23}_{-10}$ & $34^{+11}_{-5}$   & $24^{+8}_{-3}$ &  $8.5\pm0.5$   & $4.1\pm0.2$      & $2.9\pm0.2$
\enddata
\tablecomments{Assumed distances are given in Table~\ref{tab:obs}.}
\label{tab:rin_diskbb}
\end{deluxetable*}

Comparing the inner disk radii from the continuum (\verb|diskbb|) and Fe line fitting, we find that the the continuum inner disk radius is generally smaller than the Fe line value, though there are a small number of cases where the values are consistent (when assuming a mass of 1.4 M$_\odot$).  It is important to note that we have not used any spectral hardness factor or a correction for the inner boundary condition for the continuum radii.  These two factors are expected to lead to a factor of 1.18 -- 1.64 increase between the radius measured by \verb|diskbb| and the real inner disk radius \citep[e.g.][]{kubota01}.  This would bring more of the values to be consistent, though a large number would still disagree.  Where there are large disagreements, it could be possible that the continuum model chosen is not physical.  For disagreements where the continuum radius is smaller than the Fe line radius, one can imagine that disk photons may be scattered out of the line of sight by a corona, diminishing the strength of the corona, and leading to a smaller measured continuum radius.  Additionally, the continuum method relies on knowing the absolute flux of the disk component.  Therefore, if the absolute flux calibration of the telescope, or if the continuum model (column density or other parameters) are incorrect, then this can lead to an incorrect inner disk radius.

\subsection{Source inclination and disk emissivity}

\begin{figure}
\centering
\includegraphics[width=7.5cm]{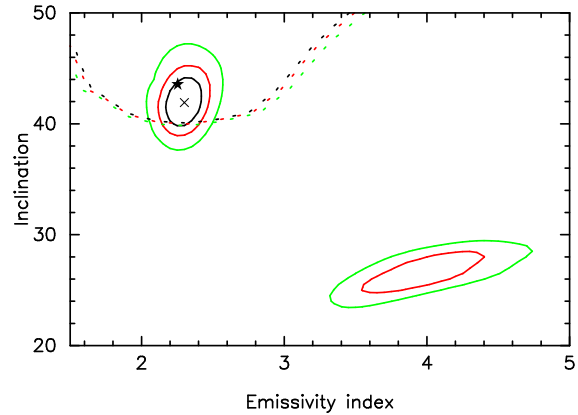}
\caption{Dependence of the $\chi^2$ confidence contours on inclination and emissivity index for GX~349+2.  The cross marks the global $\chi^2$ minimum when the second \suz{} observation is fit on its own, and the star marks the minimum when all three observations are GX~349+2 are fit jointly, with the inclination tied between them. The contours are the 1$\sigma$ (black), 90\% (red), and 99\% (green) confidence levels, with the solid lines from fitting the second observation on its own, and the dotted lines from when all three observations are fit jointly.}
\label{fig:contours}
\end{figure}

We found in many cases that there are several local $\chi^2$ minima within the \verb|diskline| parameter space.  There is usually one solution that is clearly the global minimum, with a significantly better $\chi^2$ value. However, we found that in one instance (GX~349+2, second \suz{} observation) the difference was as small as $\Delta \chi^2 \sim 2$.  This issue is due to a degeneracy where a model with a combination of low inclination, high emissivity index and high line energy can look similar to a higher inclination, lower emissivity index, and lower line energy model. Usually the inner disk radius remains consistent between the solutions.  This is apparent in Tab.~\ref{tab:phenom_line}, see particularly the GX~349+2 and Ser~X-1 observations.  We demonstrate this issue by showing the dependence of the $\chi^2$ confidence contours on inclination and emissivity index for  the second GX~349+2 \suz{} observation in Fig.~\ref{fig:contours}.  Here, one clearly sees two minima with only a small difference in $\chi^2$.  The $\chi^2$ minimum is 1262.6 for the low emissivity index, high inclination minimum compared to 1265.0 for the high emissivity index, low inclination local minimum (both have 1106 degrees of freedom).  For the first \suz{} GX~349+2 we also find two minima.  In this case, however, the global minimum has high emissivity index, low inclination opposite to the second observation ($\chi^2$ = 1268.1 for low inclination compared to 1273.6 for higher inclination).

\begin{figure}
\centering
\includegraphics[width=7.5cm]{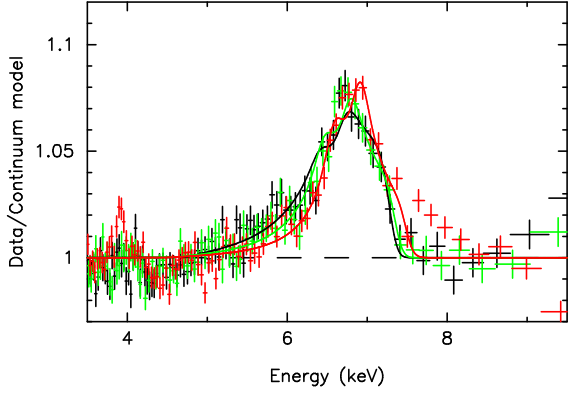}
\caption{Fe K line profile (shown as data/continuum model) for the three observations of GX~349+2.  Here the inclination was tied between the observations during the spectral fitting.  Black shows the first \suz{} observation, green the second, and red the only \xmm{} observation.}
\label{fig:gx349_all}
\end{figure}

As discussed, comparing all observations of GX~349+2, the inclination and emissivity index from spectral fitting varies significantly (see Tab.~\ref{tab:phenom_line}).  This was first pointed out by \citet{iaria09} when comparing their \xmm{} results with the \citet{cackett08} results.  As they also noted, while it is possible that emissivity index and inner disk radius and the line energy may vary with source state, the source inclination obviously does not.  As we have demonstrated, this difference is not because the lines look very different.  In fact, if all three observations of GX~349+2 are fit simultaneously with the inclination tied between observations (best fit value $i = 44^{+11}_{-4}$ degrees), then a good fit is achieved, and the lines look almost identical (see Fig.~\ref{fig:gx349_all}).  Plainly, a priori knowledge of source inclination is of significant benefit.  However, joint fitting of multiple observations of the same source potentially resolves this issue.  As can been seen in Fig.~\ref{fig:contours} when the three observations of GX~349+2 are fit jointly, there is no longer a local $\chi^2$ minimum with a high emissivity index and low inclination.

Such a degeneracy between the line parameters is self-evident when studying how each parameter affects the line profile \citep[see fig. 1 of][]{fabian89}. The inclination shifts the peak of the blue wing of the line profile.  Therefore, a higher line energy can be offset by a lower inclination.  Moreover, a higher inclination makes the line broader and less `peaked', and the emissivity index also has a similar effect.  Thus, a higher inclination, and lower emissivity index can look similar to a lower inclination and higher emissivity index.

We find a similar problem with the inclination determined from fitting the Cyg~X-2 spectra, where we measure an inclination of $25^{\circ}$. This inclination is not consistent with the value measured from optical observations, where lower and upper inclination limits are $49^\circ$ -- $73^\circ$, with a best-fitting value of $63^\circ$ \citep{orosz99,elebert09}. If we constrain the inclination to be within the optical values we do not achieve as good a fit ($\chi^2 = 1337$ compared to $\chi^2 = 1305$), and find the inclination pegs at the lower bound of $49^\circ$.  With the inclination constrained, the line parameters we find are $E = 6.40^{+0.01}$ keV, $q = 2.9\pm0.2$, $R_{\rm in} = 25\pm4$ $GM/c^2$, $i = 49^{+1\circ}$. Again, there is a pattern where a higher inclination leads to a lower line energy and emissivity index.

Note that in several cases we find high emissivity indices and/or very low inclinations when fitting the reflection models.  Both a high emissivity index and low inclination try to make to line profile more peaked.  We tested whether this can be solved with a higher Fe abundance by using a reflection model with two times solar abundance, but retrieved similar results, and worse fits.  Light bending can potentially lead to a high emissivity index, causing the irradiation of the disk to be more centrally concentrated. In the most extreme case of spinning black holes an index of $\sim$ 5 is expected \citep{miniutti03}.  These effects should not be as strong in neutron stars, and thus a lower index would be expected.

It is also important to note that in the reflection modeling the inclination is only introduced by the relativistic blurring, and the reflection spectrum itself is angle-averaged.  However, reflection is sensitive to angle, with the reflection strength weakening with higher inclination.  The use of an angle-averaged model may therefore bias the results obtained.

In black hole systems, the ionizing flux is dominated by a power-law.  The
Compton back-scattering part of the reflection spectrum is easily detected
against a power-law continuum.  Disk reflection is more difficult to detect
against a thermal spectrum since the continuum flux drops precipitously
with higher energy.  In the case of neutron stars, then, disk reflection
modeling may only yield improved physical constraints in the limit of
extraordinary high energy sensitivity.  The fact that many of our
reflection fits imply small radii and low inclinations may be partially due
to a combination of modest high energy sensitivity, coupled with the
limitations of angle-averaged disk reflection models.  Deeper observations
with \suz{} aimed at clearly detecting the Compton back-scattering hump
will help to obtain better physical constraints.

\subsection{Ionization parameter and reflection fraction}

Reflection modeling allows us to infer the ionization state of the inner accretion disk.  The reflection spectrum is highly dependent on the ionization parameter, $\xi$ \citep[see fig. 1 of][]{ballantyne04}. 
For low ionization states many emission lines are present which disappear as the ions become more highly ionized.  The Fe K$\alpha$ line in particular, evolves significantly with ionization parameter.  For neutral reflection, the 6.4 keV line is strong and as the ionization increases this line becomes weaker. The 6.7 keV line then becomes prominent until it disappears as iron becomes fully ionized.  This evolution is displayed clearly in figure 3 of \citet{ballantyne04}, where the dependence of the equivalent width (EW) of the Fe K$\alpha$ on ionization parameter is shown.  The EW peaks for either neutral reflection when the 6.4 keV line strong, or for $\log{\xi} = 2.6$ -- 2.8 when the 6.7 keV line is dominant.  From our reflection fitting (Tables \ref{tab:reflect_cont} and \ref{tab:reflect_line}) we find quite a narrow range of ionization parameter.  Most spectra are best fit with an ionization parameter in the range $\log{\xi} = 2.6$ -- 2.8, where the Fe K$\alpha$ EW peaks.

We also consider the strength of the reflection component, which directly relates to the solid angle subtended by the reflector, $\Omega$.  The reflection fraction, $R = \Omega/2\pi$, is given in Table \ref{tab:reflect_line}.  A reflection fraction of 1 corresponds an isotropic source above the slab.  Most of the observations here have a reflection fraction in the range $R =$ 0.1 -- 0.3, suggesting that a reasonable fraction of the ionizing flux is intercepted by the accretion disk and re-emitted.  If we interpret the blackbody component as the boundary layer, this may also suggest that the boundary layer has quite a similar extent in these sources.  As discussed above, figure 3 of \citet{ballantyne04} shows a strong dependence of the Fe K$\alpha$ line EW and the blackbody temperature.  Most of the blackbody temperatures we find have $kT > 2$ keV, which equates to an EW of $> 200$ eV for $R=1$.  From the phenomenological fits (see Table~\ref{tab:phenom_line}), we get equivalent widths that are mostly less than 100 eV, thus, the reflection fractions that we find generally seem reasonable.

There are two cases (one spectra of 4U 1705$-$44 and one spectra of 4U 1636$-$53) where the best-fitting reflection fractions are at the upper limit of the model ($R = 5$). In these spectra, either the blackbody temperature or the blackbody normalization is very low.  Both these lead to a low equivalent width, and therefore for the model to achieve the required line strength a large reflection fraction is required.  Both these sources are atolls which do transit between hard and soft states \citep[see, e.g.,][]{gladstone07}, and these two particular observations are in the hard state.  As discussed earlier, spectral deconvolution in atoll and Z sources is not unique \citep[e.g.][]{lin07}, and the high values of the reflection fraction suggest that our particular choice of model here is unlikely the best physical model.  We have also tried replacing the blackbody reflection model with a power-law reflection model.  While this achieved more reasonable reflection fractions it did not provide as good a fit to the data.  \citet{dai10} have recently successfully modelled the hard state in 4U 1705-44 using an alternative model involving a Comptonization model to fit the hard component, and such a model may be more appropriate for these two particular observations, though we do not investigate this here.

\subsection{The boundary layer illuminating the disk}

It is well known that a hot blackbody (or alternatively a Comptonized component) dominates from around 7 -- 20 keV \citep[e.g.][]{barret00}.  In all sources but the accreting millisecond pulsars, our spectral fitting agrees with this.  This blackbody component may potentially be due to emission from the boundary layer between the inner accretion disk and neutron star surface \citep[see, e.g., ][for discussions about the boundary layer]{sunyaev86,inogamov99,popham01,grebenev02,gilfanov03,revnivtsev06}.  The boundary layer is the transition region between the rapidly spinning accretion disk and the more slowly spinning neutron star, and should account for a significant fraction of the accretion luminosity.

As neutron star spectra show that the blackbody component dominates the flux able to ionize iron, it suggests that the boundary layer may be the dominant source of ionizing flux illuminating the inner accretion disk, leading to Fe K$\alpha$ emission.   Our spectral analysis with reflection models shows that this picture is consistent with the data.

It is not a new idea that the boundary layer or neutron star surface can illuminate the disk leading to Fe K emission. For instance, \citet{daydone91} were the first to note that X-ray bursts on the surface of the neutron star can lead to a disk-reflection component.  Furthermore, \citet{brandt94} use a geometry where an inner spherical cloud of Comptonized gas acts to illuminate the inner accretion disk as a basis for calculating Fe K profiles expected in Z sources.  Moreover, detailed calculations of the inner disk and boundary layer led \citet{popham01} to conclude that the X-ray flux from the hot boundary layer incident on the accretion disk would be large enough to lead to line emission and a reflection spectrum.

Observationally, there is not a large body of previous work fitting broadband neutron star spectra with reflection models where the boundary layer irradiates the accretion disk. However,  \citet{done02} successfully fit {\it Ginga} data of Cyg~X-2 with a reflection model which illuminates the disk with a Comptonized continuum, and similarly \citet{gierlinski02} achieved good fits to the {\it RXTE} spectra of atoll 4U~1608$-$52 with the same model.  However, the data in both cases had significantly lower spectral resolution to those presented here.  \citet{disalvo00_1728} also successfully fit the broadband {\it BeppoSAX} spectrum of MXB~1728$-$34 with a reflection model which illuminates the disk with a Comptonized continuum.  Furthermore, \citet{barret00} suggest that the boundary layer provides the X-ray flux irradiating the disk from spectral fitting of {\it RXTE} data.

The \citet{popham01} model for the boundary layer is one where the hot boundary layer gas is radially and vertically extended. \citet{popham01} showed that the hot boundary layer is much thicker than the inner disk (see their fig. 1).  As mass accretion rate changes they find that both the radial and vertical extent of the boundary layer will change, and thus the fraction of X-ray emission from the boundary layer that is intercepted by the disk will vary.  An alternative model for the boundary layer is one where matter spreads over the neutron star surface in a latitudinal belt \citep{inogamov99}.  In this spreading layer the belt width is dependent on the mass accretion rate.  At low mass accretion rates ($3\times10^{-3}$) the width is of the order of the thin-disk thickness, however, the belt disappears at high Eddington fractions ($\sim$0.9) as the entire surface is radiating.  It therefore seems reasonable that even the spreading layer may plausibly illuminate the inner accretion disk.

We now further investigate the notion that the boundary layer is the source of ionizing flux. Using ionization parameters typical to what we find from spectral fitting, we determine the maximum height of the boundary layer, for a disk extending close to the neutron star surface. The ionization parameter at the inner accretion disk, $\xi$, is given by
\begin{equation}
\xi = \frac{L_{\rm BL}}{nR^2}
\end{equation}
where $L_{\rm BL}$ is the boundary layer luminosity, n is the number density in the disk, and $R$ is the distance from the ionizing source to the disk.  For the purposes of this estimation we use the distance between the top of the boundary layer and the inner accretion disk.  That distance is given by
\begin{equation}
R = \sqrt{R_{\rm in}^2 + Z^2}
\end{equation}
where $R_{\rm in}$ is the inner disk radius, and $Z$ is the height of the ionizing source above the disk.  The distance will, of course, be higher for other regions of the accretion disk, and for boundary layer emission from closer to the equator.  However, combining these two equations and re-arranging leads to the height of the ionizing source above the disk being defined as
\begin{equation}
Z^2 = \frac{L_{\rm BL}}{n\xi} - R_{\rm in}^2 \, .
\end{equation}
From spectral fitting and reflection modeling we determine $L_{\rm BL}$, $R_{\rm in}$ and $\xi$, and from standard disk theory we can put reasonable constraints on $n$.  From \citet{shakura73}, we find $n > 10^{21}$ cm$^{-3}$.  For $\log{\xi} = 3$, $L_{\rm BL} = 10^{37}$ erg s$^{-1}$ and $R_{\rm in} = 20$ km, we find $Z < 24$ km. It is therefore completely reasonable that the boundary layer provides the ionizing flux to produce the Fe K emission lines observed.  A subsequent detailed analysis of multiple spectra from 4U~1705$-$44 has also concluded that the boundary layer likely illuminates the disk, at least in high flux states \citep{dai10}.

A further check can be performed by looking at the inferred blackbody emitting radius from the spectral fits.  In Table~\ref{tab:bb} we show the calculated emitting radius and assumed distances from the blackbody normalizations.  We use the same distances as given in Table~\ref{tab:obs}. As the boundary layer emission is more correctly described by a Comptonized component as opposed to a simple blackbody we can maybe not interpret these numbers directly.  However, they do provide a good consistency check, with radii typically a significant fraction of the stellar radius, as expected theoretically \citep{popham01}.  In a couple of cases the radii are extremely high, in those few cases. it seems likely that while the model may be a good fit to the data it is not physically correct.

\begin{deluxetable*}{lclcc}
\tablecolumns{5}
\tablewidth{0pc}
\tablecaption{Blackbody emitting radius}
\tablehead{Source & Mission & Obs. ID & BB radius (km) & BB radius (km) \\
 & & & Phenomenological & Reflection }
\startdata
Serpens~X-1    & \suz & 401048010  &  $5\pm1$      &  $5\pm1$ \\
               & \xmm & 0084020401 &  $7\pm2$      &  $7\pm2$ \\
               & \xmm & 0084020501 &  $6\pm1$      &  $5\pm1$ \\
               & \xmm & 0084020601 &  $7\pm2$      &  $6\pm1$ \\
4U~1636$-$53   & \xmm & 0303250201 &  $1.5\pm0.1$  &  $1.0\pm0.2$ \\
               & \xmm & 0500350301 &  $4.0\pm0.1$  &  $3.1\pm0.5$ \\
               & \xmm & 0500350401 &  $4.5\pm0.1$  &  $3.8\pm0.3$ \\
4U~1705$-$44   & \suz & 401046010  &  $3.5\pm0.4$  &  $1.1^{+0.9}_{-0.3}$\\
               & \suz & 401046020  &  $5.2\pm0.2$  &  $3.9\pm0.4$\\
               & \suz & 401046030  &  $3.9\pm0.2$  & $3.2\pm0.2$\\
               & \xmm & 0402300201 &  $0.7\pm0.1$  & $0.7\pm0.1$\\
4U~1820$-$30   & \suz & 401047010  &  $5.4\pm0.3$  & $5.4\pm0.3$\\
GX~17+2        & \suz & 402050010  &  $3.1\pm0.1$  & $3.3\pm0.4$\\
               & \suz & 402050020  &  $4.2\pm0.2$  & $4.1\pm0.3$\\
GX~349+2       & \suz & 400003010  &  $12\pm3$     & $11\pm3$ \\
               & \suz & 400003020  &  $10\pm3$     & $10\pm2$\\
               & \xmm & 0506110101 &  $21\pm5$     & $19\pm5$\\
Cyg~X-2      & \suz & 403063010    &  $3.1\pm0.6$  & $3.0\pm0.6$\\
SAX~J1808.4$-$3658   & \suz & 903003010  &  $9.0\pm0.8$ & $6.8\pm0.7$\\
                     & \xmm & 0560180601 &  $56\pm4$ & $540\pm74$

\enddata
\tablecomments{Assumed distances are given in Table~\ref{tab:obs}.}
\label{tab:bb}
\end{deluxetable*}

In many cases, we find that the hot blackbody component contributes substantially to the overall flux.  Substantial boundary layer emission is consistent with observed spins in neutron stars in LMXBs, which spin slower than the Keplerian angular speed in the inner accretion disk.  If a neutron star had significantly higher angular speed, the boundary layer emission is expected to tend to zero as the angular speed approached the Keplerian angular speed \citep[see fig. 5 from][for example]{bhattacharyya00}.

\subsection{Alternative emission line model}

We interpret the observed Fe K emission line as originating in the
inner accretion disk, with relativistic effects giving rise to the
asymmetric line profile (as is generally accepted). There has, however,
been a suggestion that these asymmetric Fe K lines can originate from
an optically-thick high-velocity flow in these systems
\citep{titarchuk03,laming04,laurent07,shaposhnikov09,titarchuk09}. In
this model, a diverging wind with a wide opening angle is launched
from the disk.  The Fe K line is formed in a narrow wind shell that is
illuminated by X-rays from the accretion disk. The asymmetric profile
arises when Fe K photons scatter off electrons in the partly ionized
wind, becoming red-shifted in the process.  These photons undergo a
large number of multiple scatterings to form the red wing of the
line. Such a model appears to fit some Fe K line profiles well
\citep{shaposhnikov09,titarchuk09}.

To reproduce the observed line profiles requires that there be a very
large number of electron scatters in the wind, and hence the wind had
a high optical depth \citep[][find $\tau_{\rm wind} = 1.6$ for neutron
star LMXB Ser~X-1 and $\tau_{\rm wind} = 4.9$ for black hole X-ray
binary GX~339$-$4]{titarchuk09}.  The mass outflow rate required for
GX~339$-$4 is comparable to the Eddington mass inflow rate.  In the
low/hard state considered by \citet{titarchuk09}, this means
that the outflow rate is 30 -- 100 times higher than the inferred mass
inflow rate.  The outflow rate required to produce the line in
Ser~X-1 is also comparable to the Eddington mass inflow rate.  Not only are high outflow rates required, but the outflow must also not be coupled to the inflow rate given that we see similar line profiles in the neutron star systems over more than two orders of magnitude in luminosity.

Wide-angle, optically-thick outflows should have observable
consequences in X-rays and in other bands.  In X-rays, ionizing flux from the central engine should generate absorption lines for any line of sight through the wind.  Yet when absorption lines are observed in black holes, they are always slow, with velocities less than approximately 1000~km/s (see, e.g., \citealt{miller06_nature, miller08_j1655} on GRO~J1655$-$40 and \citealt{ueda09} on GRS~1915$+$105).  In some cases, absorption lines are seen in the
absence of any relativistic line \citep[e.g. H 1743$-$322 and 4U 1630$-$472,][]{miller06_H1743,kubota07}.  The absorption lines
detected in most neutron star systems (the edge-on ``dipping" sources)
are often consistent with no velocity shift
\citep[e.g.][]{sidoli01,parmar02,boirin05,diaztrigo06}.  Clearly,
observed absorption spectra are in strong disagreement with the wind
model.

In optical and IR bands, optically-thick diverging outflows should
also have consequences.  Whether in a black hole, neutron star, or
white dwarf system, the outflow should serve to obscure our view of
the outer accretion disk and/or companion star, at least during some
intervals of the binary period.  Along particular lines of sight, the
outer disk should be obscured for nearly all of the binary period.
Obscuration of this kind has not been reported, though it would surely
be easily detected.  Doppler tomography reveals no
evidence of such outflows as accretion disk, stream/hot spot and companion star emission are often clearly revealed \citep[e.g.][]{marsh94,steeghs02}.

Finally, it is worth noting that relativistic lines are not detected
in white dwarf systems.  \citet{titarchuk09} fit their model to the
\xmm{} CCD spectrum of GK Per, where they only observe one emission
line. This, however, ignores the fact that high-resolution {\it
Chandra} gratings observations clearly separated the emission into 3
separate Fe components, as is seen in many magnetic CVs
\citep{hellier04}.  The spectra of CVs is best described in terms of
partially obscured emission from the boundary layer, which can be
fitted with cooling-flow models that predict multiple iron charge
states \citep{hellier04}.

\subsection{Equation of State}

\begin{figure}
\centering
\includegraphics[width=6cm]{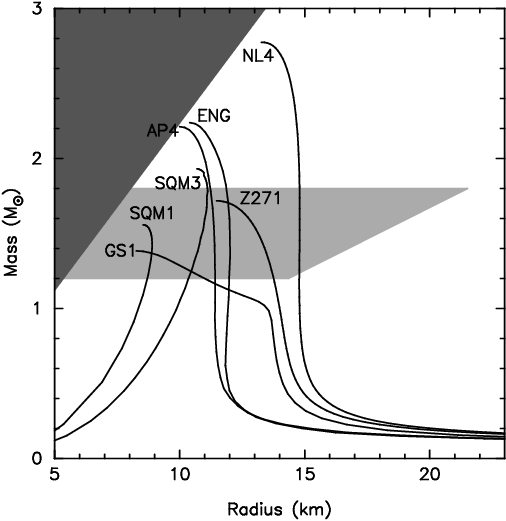}
\caption{Constraints on the ultra-dense matter equation of state from the mass and radius upper limit of Cyg~X-2 (light gray region).  The dark gray region is not allowed due to causality.   All mass-radius curves are as labeled in \citet{latt_prak01} (see references therein for details of the equations of state), except for NL4 which is from \citet{akmal98} and Z271 which is from  \citet{horowitz01}. }
\label{fig:eos}
\end{figure}

To begin to put meaningful constraints on the neutron star equation of state, it is useful to combine both mass and radius measurements.  Of the sources in our sample, Cyg~X-2 is the only one with a well constrained mass measurement, $M_{\rm NS} = 1.5 \pm 0.3 M_\odot$ \citep{elebert09}.  Combining this mass measurement with our stellar radius upper limit (8.1 $GM/c^2$) can be seen in Fig.~\ref{fig:eos}.  This does not rule out any of the possible equations of state shown here, and tighter constraints on mass and radius would be needed.

Significantly improved sensitivity and spectral resolution from future X-ray missions, such as {\it Astro-H} and the {\it International X-ray Observatory (IXO)} will certainly provide tighter constraints from modeling Fe K emission lines, and with the High Timing Resolution Spectrometer on {\it IXO} even the potential to study simultaneous spectral and timing evolution on extremely short timescales \citep{barret08}.

\acknowledgements We appreciate the extremely helpful comments from the referee which have improved the paper.  EMC gratefully acknowledges support provided by NASA
through the Chandra Fellowship Program, grant number PF8-90052. This research has made use of data obtained from the \suz{} satellite, a collaborative mission between the space agencies of Japan (JAXA) and the USA (NASA).  It also made use of observations obtained with \xmm, an ESA science mission with instruments and contributions directly funded by ESA Member States and NASA.

\bibliographystyle{apj}
\bibliography{apj-jour,feline}

\begin{thebibliography}{115}
\expandafter\ifx\csname natexlab\endcsname\relax\def\natexlab#1{#1}\fi

\bibitem[{{Akmal} {et~al.}(1998){Akmal}, {Pandharipande}, \&
  {Ravenhall}}]{akmal98}
{Akmal}, A., {Pandharipande}, V.~R., \& {Ravenhall}, D.~G. 1998, \prc, 58, 1804

\bibitem[{{Altamirano} {et~al.}(2010){Altamirano}, {Hiemstra}, {M{\`e}ndez},
  {Homan}, \& {Belloni}}]{altamirano09}
{Altamirano}, D., {Hiemstra}, B., {M{\`e}ndez}, M., {Homan}, J., \& {Belloni},
  T.~M. 2010, MNRAS, submitted

\bibitem[{{Arnaud}(1996)}]{arnaud96}
{Arnaud}, K.~A. 1996, in Astronomical Society of the Pacific Conference Series,
  Vol. 101, Astronomical Data Analysis Software and Systems V, ed. G.~H.
  {Jacoby} \& J.~{Barnes}, 17

\bibitem[{{Ballantyne}(2004)}]{ballantyne04}
{Ballantyne}, D.~R. 2004, \mnras, 351, 57

\bibitem[{{Ballantyne} {et~al.}(2001){Ballantyne}, {Ross}, \&
  {Fabian}}]{ballantyne01}
{Ballantyne}, D.~R., {Ross}, R.~R., \& {Fabian}, A.~C. 2001, \mnras, 327, 10

\bibitem[{{Ballantyne} \& {Strohmayer}(2004)}]{ball_stroh04}
{Ballantyne}, D.~R., \& {Strohmayer}, T.~E. 2004, \apjl, 602, L105

\bibitem[{{Barret}(2001)}]{barret01}
{Barret}, D. 2001, Advances in Space Research, 28, 307

\bibitem[{{Barret} {et~al.}(2008){Barret}, {Belloni}, {Bhattacharyya},
  {Gilfanov}, {Gogus}, {Homan}, {M{\'e}ndez}, {Miller}, {Miller}, {Mereghetti},
  {Paltani}, {Poutanen}, {Wilms}, \& {Zdziarski}}]{barret08}
{Barret}, D. {et~al.} 2008, in Society of Photo-Optical Instrumentation
  Engineers (SPIE) Conference Series, Vol. 7011, Society of Photo-Optical
  Instrumentation Engineers (SPIE) Conference Series

\bibitem[{{Barret} {et~al.}(2000){Barret}, {Olive}, {Boirin}, {Done},
  {Skinner}, \& {Grindlay}}]{barret00}
{Barret}, D., {Olive}, J.~F., {Boirin}, L., {Done}, C., {Skinner}, G.~K., \&
  {Grindlay}, J.~E. 2000, \apj, 533, 329

\bibitem[{{Barret} {et~al.}(2005){Barret}, {Olive}, \& {Miller}}]{barret05}
{Barret}, D., {Olive}, J.-F., \& {Miller}, M.~C. 2005, \mnras, 361, 855

\bibitem[{{Barret} {et~al.}(2006){Barret}, {Olive}, \& {Miller}}]{barret06}
---. 2006, \mnras, 370, 1140

\bibitem[{{Belloni} {et~al.}(2005){Belloni}, {M{\'e}ndez}, \&
  {Homan}}]{belloni05}
{Belloni}, T., {M{\'e}ndez}, M., \& {Homan}, J. 2005, \aap, 437, 209

\bibitem[{{Bhattacharyya} \& {Strohmayer}(2007)}]{bhattacharyya07}
{Bhattacharyya}, S., \& {Strohmayer}, T.~E. 2007, \apjl, 664, L103

\bibitem[{{Bhattacharyya} {et~al.}(2000){Bhattacharyya}, {Thampan}, {Misra}, \&
  {Datta}}]{bhattacharyya00}
{Bhattacharyya}, S., {Thampan}, A.~V., {Misra}, R., \& {Datta}, B. 2000, \apj,
  542, 473

\bibitem[{{Boirin} {et~al.}(2000){Boirin}, {Barret}, {Olive}, {Bloser}, \&
  {Grindlay}}]{boirin00}
{Boirin}, L., {Barret}, D., {Olive}, J.~F., {Bloser}, P.~F., \& {Grindlay},
  J.~E. 2000, \aap, 361, 121

\bibitem[{{Boirin} {et~al.}(2005){Boirin}, {M{\'e}ndez}, {D{\'{\i}}az Trigo},
  {Parmar}, \& {Kaastra}}]{boirin05}
{Boirin}, L., {M{\'e}ndez}, M., {D{\'{\i}}az Trigo}, M., {Parmar}, A.~N., \&
  {Kaastra}, J.~S. 2005, \aap, 436, 195

\bibitem[{{Boutelier} {et~al.}(2010){Boutelier}, {Barret}, {Lin}, \&
  {T{\"o}r{\"o}k}}]{boutelier10}
{Boutelier}, M., {Barret}, D., {Lin}, Y., \& {T{\"o}r{\"o}k}, G. 2010, \mnras,
  401, 1290

\bibitem[{{Boutelier} {et~al.}(2009){Boutelier}, {Barret}, \&
  {Miller}}]{boutelier09}
{Boutelier}, M., {Barret}, D., \& {Miller}, M.~C. 2009, \mnras, 399, 1901

\bibitem[{{Brandt} \& {Matt}(1994)}]{brandt94}
{Brandt}, W.~M., \& {Matt}, G. 1994, \mnras, 268, 1051

\bibitem[{{Brenneman} \& {Reynolds}(2006)}]{brenneman06}
{Brenneman}, L.~W., \& {Reynolds}, C.~S. 2006, \apj, 652, 1028

\bibitem[{{Cackett} {et~al.}(2009{\natexlab{a}}){Cackett}, {Altamirano},
  {Patruno}, {Miller}, {Reynolds}, {Linares}, \& {Wijnands}}]{cackett_j1808_09}
{Cackett}, E.~M., {Altamirano}, D., {Patruno}, A., {Miller}, J.~M., {Reynolds},
  M., {Linares}, M., \& {Wijnands}, R. 2009{\natexlab{a}}, \apjl, 694, L21

\bibitem[{{Cackett} {et~al.}(2008){Cackett}, {Miller}, {Bhattacharyya},
  {Grindlay}, {Homan}, {van der Klis}, {Miller}, {Strohmayer}, \&
  {Wijnands}}]{cackett08}
{Cackett}, E.~M. {et~al.} 2008, \apj, 674, 415

\bibitem[{{Cackett} {et~al.}(2009{\natexlab{b}}){Cackett}, {Miller}, {Homan},
  {van der Klis}, {Lewin}, {M{\'e}ndez}, {Raymond}, {Steeghs}, \&
  {Wijnands}}]{cackett_chazss_09}
---. 2009{\natexlab{b}}, \apj, 690, 1847

\bibitem[{{Casares} {et~al.}(2006){Casares}, {Cornelisse}, {Steeghs},
  {Charles}, {Hynes}, {O'Brien}, \& {Strohmayer}}]{casares06}
{Casares}, J., {Cornelisse}, R., {Steeghs}, D., {Charles}, P.~A., {Hynes},
  R.~I., {O'Brien}, K., \& {Strohmayer}, T.~E. 2006, \mnras, 373, 1235

\bibitem[{{Christian} \& {Swank}(1997)}]{christian97}
{Christian}, D.~J., \& {Swank}, J.~H. 1997, \apjs, 109, 177

\bibitem[{{D'A{\`i}} {et~al.}(2010){D'A{\`i}}, {Di Salvo}, {Ballantyne},
  {Iaria}, {Robba}, {Papitto}, {Riggio}, {Burderi}, {Piraino}, {Santangelo},
  {Matt}, {Dov{\v c}iak}, \& {Karas}}]{dai10}
{D'A{\`i}}, A. {et~al.} 2010, A\&A, in press, arXiv:1004.1963

\bibitem[{{D'A{\`i}} {et~al.}(2009){D'A{\`i}}, {Iaria}, {Di Salvo}, {Matt}, \&
  {Robba}}]{dai09}
{D'A{\`i}}, A., {Iaria}, R., {Di Salvo}, T., {Matt}, G., \& {Robba}, N.~R.
  2009, \apjl, 693, L1

\bibitem[{{Davis}(2001)}]{davis01}
{Davis}, J.~E. 2001, \apj, 562, 575

\bibitem[{{Day} \& {Done}(1991)}]{daydone91}
{Day}, C.~S.~R., \& {Done}, C. 1991, \mnras, 253, 35P

\bibitem[{{Di Salvo} {et~al.}(2009){Di Salvo}, {D'Ai'}, {Iaria}, {Burderi},
  {Dov{\v c}iak}, {Karas}, {Matt}, {Papitto}, {Piraino}, {Riggio}, {Robba}, \&
  {Santangelo}}]{disalvo09}
{Di Salvo}, T. {et~al.} 2009, ArXiv e-prints

\bibitem[{{Di Salvo} {et~al.}(2000{\natexlab{a}}){Di Salvo}, {Iaria},
  {Burderi}, \& {Robba}}]{disalvo00_1728}
{Di Salvo}, T., {Iaria}, R., {Burderi}, L., \& {Robba}, N.~R.
  2000{\natexlab{a}}, \apj, 542, 1034

\bibitem[{{Di Salvo} {et~al.}(2000{\natexlab{b}}){Di Salvo}, {Stella}, {Robba},
  {van der Klis}, {Burderi}, {Israel}, {Homan}, {Campana}, {Frontera}, \&
  {Parmar}}]{disalvo00}
{Di Salvo}, T. {et~al.} 2000{\natexlab{b}}, \apjl, 544, L119

\bibitem[{{D{\'{\i}}az Trigo} {et~al.}(2006){D{\'{\i}}az Trigo}, {Parmar},
  {Boirin}, {M{\'e}ndez}, \& {Kaastra}}]{diaztrigo06}
{D{\'{\i}}az Trigo}, M., {Parmar}, A.~N., {Boirin}, L., {M{\'e}ndez}, M., \&
  {Kaastra}, J.~S. 2006, \aap, 445, 179

\bibitem[{{D{\'{\i}}az Trigo} {et~al.}(2007){D{\'{\i}}az Trigo}, {Parmar},
  {Miller}, {Kuulkers}, \& {Caballero-Garc{\'{\i}}a}}]{diaztrigo07}
{D{\'{\i}}az Trigo}, M., {Parmar}, A.~N., {Miller}, J., {Kuulkers}, E., \&
  {Caballero-Garc{\'{\i}}a}, M.~D. 2007, \aap, 462, 657

\bibitem[{{Done} {et~al.}(2002){Done}, {{\.Z}ycki}, \& {Smith}}]{done02}
{Done}, C., {{\.Z}ycki}, P.~T., \& {Smith}, D.~A. 2002, \mnras, 331, 453

\bibitem[{{Elebert} {et~al.}(2009){Elebert}, {Callanan}, {Torres}, \&
  {Garcia}}]{elebert09}
{Elebert}, P., {Callanan}, P.~J., {Torres}, M.~A.~P., \& {Garcia}, M.~R. 2009,
  \mnras, 395, 2029

\bibitem[{{Fabian} {et~al.}(1989){Fabian}, {Rees}, {Stella}, \&
  {White}}]{fabian89}
{Fabian}, A.~C., {Rees}, M.~J., {Stella}, L., \& {White}, N.~E. 1989, \mnras,
  238, 729

\bibitem[{{Fabian} {et~al.}(2009){Fabian}, {Zoghbi}, {Ross}, {Uttley}, {Gallo},
  {Brandt}, {Blustin}, {Boller}, {Caballero-Garcia}, {Larsson}, {Miller},
  {Miniutti}, {Ponti}, {Reis}, {Reynolds}, {Tanaka}, \& {Young}}]{fabian09}
{Fabian}, A.~C. {et~al.} 2009, \nat, 459, 540

\bibitem[{{Farinelli} {et~al.}(2005){Farinelli}, {Frontera}, {Zdziarski},
  {Stella}, {Zhang}, {van der Klis}, {Masetti}, \& {Amati}}]{farinelli05}
{Farinelli}, R., {Frontera}, F., {Zdziarski}, A.~A., {Stella}, L., {Zhang},
  S.~N., {van der Klis}, M., {Masetti}, N., \& {Amati}, L. 2005, \aap, 434, 25

\bibitem[{{Fender} \& {Hendry}(2000)}]{fender00}
{Fender}, R.~P., \& {Hendry}, M.~A. 2000, \mnras, 317, 1

\bibitem[{{Galloway} \& {Cumming}(2006)}]{galloway06}
{Galloway}, D.~K., \& {Cumming}, A. 2006, \apj, 652, 559

\bibitem[{{Galloway} {et~al.}(2008){Galloway}, {Muno}, {Hartman}, {Psaltis}, \&
  {Chakrabarty}}]{galloway08}
{Galloway}, D.~K., {Muno}, M.~P., {Hartman}, J.~M., {Psaltis}, D., \&
  {Chakrabarty}, D. 2008, \apjs, 179, 360

\bibitem[{{George} \& {Fabian}(1991)}]{george91}
{George}, I.~M., \& {Fabian}, A.~C. 1991, \mnras, 249, 352

\bibitem[{{Gierli{\'n}ski} \& {Done}(2002)}]{gierlinski02}
{Gierli{\'n}ski}, M., \& {Done}, C. 2002, \mnras, 337, 1373

\bibitem[{{Gierli{\'n}ski} {et~al.}(1999){Gierli{\'n}ski}, {Zdziarski},
  {Poutanen}, {Coppi}, {Ebisawa}, \& {Johnson}}]{gierlinski99}
{Gierli{\'n}ski}, M., {Zdziarski}, A.~A., {Poutanen}, J., {Coppi}, P.~S.,
  {Ebisawa}, K., \& {Johnson}, W.~N. 1999, \mnras, 309, 496

\bibitem[{{Gilfanov} {et~al.}(2003){Gilfanov}, {Revnivtsev}, \&
  {Molkov}}]{gilfanov03}
{Gilfanov}, M., {Revnivtsev}, M., \& {Molkov}, S. 2003, \aap, 410, 217

\bibitem[{{Gladstone} {et~al.}(2007){Gladstone}, {Done}, \&
  {Gierli{\'n}ski}}]{gladstone07}
{Gladstone}, J., {Done}, C., \& {Gierli{\'n}ski}, M. 2007, \mnras, 378, 13

\bibitem[{{Grebenev} \& {Sunyaev}(2002)}]{grebenev02}
{Grebenev}, S.~A., \& {Sunyaev}, R.~A. 2002, Astronomy Letters, 28, 150

\bibitem[{{Hellier} \& {Mukai}(2004)}]{hellier04}
{Hellier}, C., \& {Mukai}, K. 2004, \mnras, 352, 1037

\bibitem[{{Homan} {et~al.}(2002){Homan}, {van der Klis}, {Jonker}, {Wijnands},
  {Kuulkers}, {M{\'e}ndez}, \& {Lewin}}]{homan02}
{Homan}, J., {van der Klis}, M., {Jonker}, P.~G., {Wijnands}, R., {Kuulkers},
  E., {M{\'e}ndez}, M., \& {Lewin}, W.~H.~G. 2002, \apj, 568, 878

\bibitem[{{Horowitz} \& {Piekarewicz}(2001)}]{horowitz01}
{Horowitz}, C.~J., \& {Piekarewicz}, J. 2001, Physical Review Letters, 86, 5647

\bibitem[{{Iaria} {et~al.}(2009){Iaria}, {D'A{\'{\i}}}, {Di Salvo}, {Robba},
  {Riggio}, {Papitto}, \& {Burderi}}]{iaria09}
{Iaria}, R., {D'A{\'{\i}}}, A., {Di Salvo}, T., {Robba}, N.~R., {Riggio}, A.,
  {Papitto}, A., \& {Burderi}, L. 2009, ArXiv e-prints

\bibitem[{{Inogamov} \& {Sunyaev}(1999)}]{inogamov99}
{Inogamov}, N.~A., \& {Sunyaev}, R.~A. 1999, Astronomy Letters, 25, 269

\bibitem[{{Jansen} {et~al.}(2001){Jansen}, {Lumb}, {Altieri}, {Clavel}, {Ehle},
  {Erd}, {Gabriel}, {Guainazzi}, {Gondoin}, {Much}, {Munoz}, {Santos},
  {Schartel}, {Texier}, \& {Vacanti}}]{jansen01}
{Jansen}, F. {et~al.} 2001, \aap, 365, L1

\bibitem[{{Juett} {et~al.}(2004){Juett}, {Schulz}, \& {Chakrabarty}}]{juett04}
{Juett}, A.~M., {Schulz}, N.~S., \& {Chakrabarty}, D. 2004, \apj, 612, 308

\bibitem[{{Juett} {et~al.}(2006){Juett}, {Schulz}, {Chakrabarty}, \&
  {Gorczyca}}]{juett06}
{Juett}, A.~M., {Schulz}, N.~S., {Chakrabarty}, D., \& {Gorczyca}, T.~W. 2006,
  \apj, 648, 1066

\bibitem[{{Kubota} {et~al.}(2007){Kubota}, {Dotani}, {Cottam}, {Kotani},
  {Done}, {Ueda}, {Fabian}, {Yasuda}, {Takahashi}, {Fukazawa}, {Yamaoka},
  {Makishima}, {Yamada}, {Kohmura}, \& {Angelini}}]{kubota07}
{Kubota}, A. {et~al.} 2007, \pasj, 59, 185

\bibitem[{{Kubota} {et~al.}(2001){Kubota}, {Makishima}, \&
  {Ebisawa}}]{kubota01}
{Kubota}, A., {Makishima}, K., \& {Ebisawa}, K. 2001, \apjl, 560, L147

\bibitem[{{Kuulkers} {et~al.}(2003){Kuulkers}, {den Hartog}, {in't Zand},
  {Verbunt}, {Harris}, \& {Cocchi}}]{kuulkers03}
{Kuulkers}, E., {den Hartog}, P.~R., {in't Zand}, J.~J.~M., {Verbunt},
  F.~W.~M., {Harris}, W.~E., \& {Cocchi}, M. 2003, \aap, 399, 663

\bibitem[{{Kuulkers} {et~al.}(1997){Kuulkers}, {Parmar}, {Owens},
  {Oosterbroek}, \& {Lammers}}]{kuulkers97}
{Kuulkers}, E., {Parmar}, A.~N., {Owens}, A., {Oosterbroek}, T., \& {Lammers},
  U. 1997, \aap, 323, L29

\bibitem[{{Laming} \& {Titarchuk}(2004)}]{laming04}
{Laming}, J.~M., \& {Titarchuk}, L. 2004, \apjl, 615, L121

\bibitem[{{Laor}(1991)}]{laor91}
{Laor}, A. 1991, \apj, 376, 90

\bibitem[{{Lattimer} \& {Prakash}(2001)}]{latt_prak01}
{Lattimer}, J.~M., \& {Prakash}, M. 2001, \apj, 550, 426

\bibitem[{{Laurent} \& {Titarchuk}(2007)}]{laurent07}
{Laurent}, P., \& {Titarchuk}, L. 2007, \apj, 656, 1056

\bibitem[{{Leahy} {et~al.}(1983){Leahy}, {Darbro}, {Elsner}, {Weisskopf},
  {Kahn}, {Sutherland}, \& {Grindlay}}]{leahy83}
{Leahy}, D.~A., {Darbro}, W., {Elsner}, R.~F., {Weisskopf}, M.~C., {Kahn}, S.,
  {Sutherland}, P.~G., \& {Grindlay}, J.~E. 1983, \apj, 266, 160

\bibitem[{{Lin} {et~al.}(2007){Lin}, {Remillard}, \& {Homan}}]{lin07}
{Lin}, D., {Remillard}, R.~A., \& {Homan}, J. 2007, \apj, 667, 1073

\bibitem[{{Magdziarz} \& {Zdziarski}(1995)}]{magdziarz95}
{Magdziarz}, P., \& {Zdziarski}, A.~A. 1995, \mnras, 273, 837

\bibitem[{{Markoff} \& {Nowak}(2004)}]{markoff04}
{Markoff}, S., \& {Nowak}, M.~A. 2004, \apj, 609, 972

\bibitem[{{Markoff} {et~al.}(2005){Markoff}, {Nowak}, \& {Wilms}}]{markoff05}
{Markoff}, S., {Nowak}, M.~A., \& {Wilms}, J. 2005, \apj, 635, 1203

\bibitem[{{Marsh} {et~al.}(1994){Marsh}, {Robinson}, \& {Wood}}]{marsh94}
{Marsh}, T.~R., {Robinson}, E.~L., \& {Wood}, J.~H. 1994, \mnras, 266, 137

\bibitem[{{M{\'e}ndez}(2006)}]{mendez06}
{M{\'e}ndez}, M. 2006, \mnras, 371, 1925

\bibitem[{{Miller}(2007)}]{miller07}
{Miller}, J.~M. 2007, \araa, 45, 441

\bibitem[{{Miller} {et~al.}(2006{\natexlab{a}}){Miller}, {Homan}, {Steeghs},
  {Rupen}, {Hunstead}, {Wijnands}, {Charles}, \& {Fabian}}]{miller06}
{Miller}, J.~M., {Homan}, J., {Steeghs}, D., {Rupen}, M., {Hunstead}, R.~W.,
  {Wijnands}, R., {Charles}, P.~A., \& {Fabian}, A.~C. 2006{\natexlab{a}},
  \apj, 653, 525

\bibitem[{{Miller} {et~al.}(2006{\natexlab{b}}){Miller}, {Raymond}, {Fabian},
  {Steeghs}, {Homan}, {Reynolds}, {van der Klis}, \&
  {Wijnands}}]{miller06_nature}
{Miller}, J.~M., {Raymond}, J., {Fabian}, A., {Steeghs}, D., {Homan}, J.,
  {Reynolds}, C., {van der Klis}, M., \& {Wijnands}, R. 2006{\natexlab{b}},
  \nat, 441, 953

\bibitem[{{Miller} {et~al.}(2006{\natexlab{c}}){Miller}, {Raymond}, {Homan},
  {Fabian}, {Steeghs}, {Wijnands}, {Rupen}, {Charles}, {van der Klis}, \&
  {Lewin}}]{miller06_H1743}
{Miller}, J.~M. {et~al.} 2006{\natexlab{c}}, \apj, 646, 394

\bibitem[{{Miller} {et~al.}(2008{\natexlab{a}}){Miller}, {Raymond}, {Reynolds},
  {Fabian}, {Kallman}, \& {Homan}}]{miller08_j1655}
{Miller}, J.~M., {Raymond}, J., {Reynolds}, C.~S., {Fabian}, A.~C., {Kallman},
  T.~R., \& {Homan}, J. 2008{\natexlab{a}}, \apj, 680, 1359

\bibitem[{{Miller} {et~al.}(2008{\natexlab{b}}){Miller}, {Reynolds}, {Fabian},
  {Cackett}, {Miniutti}, {Raymond}, {Steeghs}, {Reis}, \& {Homan}}]{miller08}
{Miller}, J.~M. {et~al.} 2008{\natexlab{b}}, \apjl, 679, L113

\bibitem[{{Miller} {et~al.}(2009){Miller}, {Reynolds}, {Fabian}, {Miniutti}, \&
  {Gallo}}]{miller09}
{Miller}, J.~M., {Reynolds}, C.~S., {Fabian}, A.~C., {Miniutti}, G., \&
  {Gallo}, L.~C. 2009, \apj, 697, 900

\bibitem[{{Miller} {et~al.}(1998){Miller}, {Lamb}, \& {Cook}}]{miller98_rot}
{Miller}, M.~C., {Lamb}, F.~K., \& {Cook}, G.~B. 1998, \apj, 509, 793

\bibitem[{{Miniutti} {et~al.}(2003){Miniutti}, {Fabian}, {Goyder}, \&
  {Lasenby}}]{miniutti03}
{Miniutti}, G., {Fabian}, A.~C., {Goyder}, R., \& {Lasenby}, A.~N. 2003,
  \mnras, 344, L22

\bibitem[{{Miniutti} {et~al.}(2009)}]{miniutti09}
{Miniutti}, G., {et~al.} 2009, MNRAS, in press, arXiv:astro-ph/0905.2891

\bibitem[{{Mitsuda} {et~al.}(1989){Mitsuda}, {Inoue}, {Nakamura}, \&
  {Tanaka}}]{mitsuda89}
{Mitsuda}, K., {Inoue}, H., {Nakamura}, N., \& {Tanaka}, Y. 1989, \pasj, 41, 97

\bibitem[{{Mitsuda} {et~al.}(2007)}]{mitsuda07}
{Mitsuda}, K., {et~al.} 2007, \pasj, 59, 1

\bibitem[{{Nayakshin} \& {Kallman}(2001)}]{nayakshin01}
{Nayakshin}, S., \& {Kallman}, T.~R. 2001, \apj, 546, 406

\bibitem[{{Orosz} \& {Kuulkers}(1999)}]{orosz99}
{Orosz}, J.~A., \& {Kuulkers}, E. 1999, \mnras, 305, 132

\bibitem[{{Pandel} {et~al.}(2008){Pandel}, {Kaaret}, \& {Corbel}}]{pandel08}
{Pandel}, D., {Kaaret}, P., \& {Corbel}, S. 2008, \apj, 688, 1288

\bibitem[{{Papitto} {et~al.}(2009){Papitto}, {Di Salvo}, {D'A{\`i}}, {Iaria},
  {Burderi}, {Riggio}, {Menna}, \& {Robba}}]{papitto09}
{Papitto}, A., {Di Salvo}, T., {D'A{\`i}}, A., {Iaria}, R., {Burderi}, L.,
  {Riggio}, A., {Menna}, M.~T., \& {Robba}, N.~R. 2009, \aap, 493, L39

\bibitem[{{Parmar} {et~al.}(2002){Parmar}, {Oosterbroek}, {Boirin}, \&
  {Lumb}}]{parmar02}
{Parmar}, A.~N., {Oosterbroek}, T., {Boirin}, L., \& {Lumb}, D. 2002, \aap,
  386, 910

\bibitem[{{Piraino} {et~al.}(2000){Piraino}, {Santangelo}, \&
  {Kaaret}}]{piraino00}
{Piraino}, S., {Santangelo}, A., \& {Kaaret}, P. 2000, \aap, 360, L35

\bibitem[{{Piraino} {et~al.}(2002){Piraino}, {Santangelo}, \&
  {Kaaret}}]{piraino02}
---. 2002, \apj, 567, 1091

\bibitem[{{Popham} \& {Sunyaev}(2001)}]{popham01}
{Popham}, R., \& {Sunyaev}, R. 2001, \apj, 547, 355

\bibitem[{{Psaltis} {et~al.}(1999){Psaltis}, {Wijnands}, {Homan}, {Jonker},
  {van der Klis}, {Miller}, {Lamb}, {Kuulkers}, {van Paradijs}, \&
  {Lewin}}]{psaltis99}
{Psaltis}, D. {et~al.} 1999, \apj, 520, 763

\bibitem[{{Reis} {et~al.}(2009{\natexlab{a}}){Reis}, {Fabian}, {Ross}, \&
  {Miller}}]{reis09_spin}
{Reis}, R.~C., {Fabian}, A.~C., {Ross}, R.~R., \& {Miller}, J.~M.
  2009{\natexlab{a}}, \mnras, 395, 1257

\bibitem[{{Reis} {et~al.}(2009{\natexlab{b}}){Reis}, {Fabian}, \&
  {Young}}]{reis09_1705}
{Reis}, R.~C., {Fabian}, A.~C., \& {Young}, A.~J. 2009{\natexlab{b}}, ArXiv
  e-prints

\bibitem[{{Revnivtsev} \& {Gilfanov}(2006)}]{revnivtsev06}
{Revnivtsev}, M.~G., \& {Gilfanov}, M.~R. 2006, \aap, 453, 253

\bibitem[{{Reynolds} {et~al.}(2009){Reynolds}, {Nowak}, {Markoff}, {Tueller},
  {Wilms}, \& {Young}}]{reynolds09}
{Reynolds}, C.~S., {Nowak}, M.~A., {Markoff}, S., {Tueller}, J., {Wilms}, J.,
  \& {Young}, A.~J. 2009, \apj, 691, 1159

\bibitem[{{Ross} \& {Fabian}(2007)}]{ross07}
{Ross}, R.~R., \& {Fabian}, A.~C. 2007, \mnras, 381, 1697

\bibitem[{{Schmoll} {et~al.}(2009){Schmoll}, {Miller}, {Volonteri}, {Cackett},
  {Reynolds}, {Fabian}, {Brenneman}, {Miniutti}, \& {Gallo}}]{schmoll09}
{Schmoll}, S. {et~al.} 2009, ApJ, in press, arXiv:0908.0013

\bibitem[{{Schulz}(1999)}]{schulz99}
{Schulz}, N.~S. 1999, \apj, 511, 304

\bibitem[{{Schulz} {et~al.}(2009){Schulz}, {Huenemoerder}, {Ji}, {Nowak},
  {Yao}, \& {Canizares}}]{schulz09}
{Schulz}, N.~S., {Huenemoerder}, D.~P., {Ji}, L., {Nowak}, M., {Yao}, Y., \&
  {Canizares}, C.~R. 2009, \apjl, 692, L80

\bibitem[{{Shakura} \& {Sunyaev}(1973)}]{shakura73}
{Shakura}, N.~I., \& {Sunyaev}, R.~A. 1973, \aap, 24, 337

\bibitem[{{Shaposhnikov} {et~al.}(2009){Shaposhnikov}, {Titarchuk}, \&
  {Laurent}}]{shaposhnikov09}
{Shaposhnikov}, N., {Titarchuk}, L., \& {Laurent}, P. 2009, \apj, 699, 1223

\bibitem[{{Sidoli} {et~al.}(2001){Sidoli}, {Oosterbroek}, {Parmar}, {Lumb}, \&
  {Erd}}]{sidoli01}
{Sidoli}, L., {Oosterbroek}, T., {Parmar}, A.~N., {Lumb}, D., \& {Erd}, C.
  2001, \aap, 379, 540

\bibitem[{{Steeghs} \& {Casares}(2002)}]{steeghs02}
{Steeghs}, D., \& {Casares}, J. 2002, \apj, 568, 273

\bibitem[{{Stella} \& {Vietri}(1998)}]{stella98}
{Stella}, L., \& {Vietri}, M. 1998, \apjl, 492, L59

\bibitem[{{Sunyaev} \& {Shakura}(1986)}]{sunyaev86}
{Sunyaev}, R.~A., \& {Shakura}, N.~I. 1986, Soviet Astronomy Letters, 12, 117

\bibitem[{{Titarchuk} {et~al.}(2003){Titarchuk}, {Kazanas}, \&
  {Becker}}]{titarchuk03}
{Titarchuk}, L., {Kazanas}, D., \& {Becker}, P.~A. 2003, \apj, 598, 411

\bibitem[{{Titarchuk} {et~al.}(2009){Titarchuk}, {Laurent}, \&
  {Shaposhnikov}}]{titarchuk09}
{Titarchuk}, L., {Laurent}, P., \& {Shaposhnikov}, N. 2009, ApJ, in press,
  arXiv:0906.1490

\bibitem[{{Ueda} {et~al.}(2009){Ueda}, {Yamaoka}, \& {Remillard}}]{ueda09}
{Ueda}, Y., {Yamaoka}, K., \& {Remillard}, R. 2009, \apj, 695, 888

\bibitem[{{van der Klis}(2006)}]{vanderklis06}
{van der Klis}, M. 2006, {Rapid X-ray Variability}, ed. W.~H.~G. {Lewin} \&
  M.~{van der Klis} (Cambridge University Press), 39--112

\bibitem[{{Vrtilek} {et~al.}(1988){Vrtilek}, {Swank}, \& {Kallman}}]{vrtilek88}
{Vrtilek}, S.~D., {Swank}, J.~H., \& {Kallman}, T.~R. 1988, \apj, 326, 186

\bibitem[{{White} {et~al.}(1988){White}, {Stella}, \& {Parmar}}]{white88}
{White}, N.~E., {Stella}, L., \& {Parmar}, A.~N. 1988, \apj, 324, 363

\bibitem[{{Wijnands} {et~al.}(1998){Wijnands}, {Homan}, {van der Klis},
  {Kuulkers}, {van Paradijs}, {Lewin}, {Lamb}, {Psaltis}, \&
  {Vaughan}}]{wijnands98}
{Wijnands}, R. {et~al.} 1998, \apjl, 493, L87

\bibitem[{{Wijnands} {et~al.}(1997){Wijnands}, {Homan}, {van der Klis},
  {Mendez}, {Kuulkers}, {van Paradijs}, {Lewin}, {Lamb}, {Psaltis}, \&
  {Vaughan}}]{wijnands97}
---. 1997, \apjl, 490, L157

\bibitem[{{Wilkinson} \& {Uttley}(2009)}]{wilkinson09}
{Wilkinson}, T., \& {Uttley}, P. 2009, \mnras, 824

\end{thebibliography}

\end{document}